\documentclass{article}

\usepackage[T1]{fontenc}  
\usepackage[utf8]{inputenc}
\usepackage[a4paper]{geometry}

\usepackage{amsmath,amsfonts,amssymb}
\usepackage{stmaryrd} 
\usepackage[pdftex]{graphicx}
\usepackage{tikz}
\usetikzlibrary{matrix}

\usepackage{authblk}
\usepackage{algorithm}
\usepackage{algpseudocode}
\usepackage{caption}
\usepackage{subcaption}
\usepackage{graphicx}
\usepackage{cleveref}

\usepackage{array,multirow,makecell}
\setcellgapes{1pt}
\makegapedcells
\newcolumntype{R}[1]{>{\raggedleft\arraybackslash }b{#1}}
\newcolumntype{L}[1]{>{\raggedright\arraybackslash }b{#1}}
\newcolumntype{C}[1]{>{\centering\arraybackslash }b{#1}}

\usepackage{lineno}
\usepackage{lineno}
\newcommand*\patchAmsMathEnvironmentForLineno[1]{%
  \expandafter\let\csname old#1\expandafter\endcsname\csname #1\endcsname
  \expandafter\let\csname oldend#1\expandafter\endcsname\csname end#1\endcsname
  \renewenvironment{#1}%
     {\linenomath\csname old#1\endcsname}%
     {\csname oldend#1\endcsname\endlinenomath}}%
\newcommand*\patchBothAmsMathEnvironmentsForLineno[1]{%
  \patchAmsMathEnvironmentForLineno{#1}%
  \patchAmsMathEnvironmentForLineno{#1*}}%
\AtBeginDocument{%
\patchBothAmsMathEnvironmentsForLineno{equation}%
\patchBothAmsMathEnvironmentsForLineno{align}%
\patchBothAmsMathEnvironmentsForLineno{flalign}%
\patchBothAmsMathEnvironmentsForLineno{alignat}%
\patchBothAmsMathEnvironmentsForLineno{gather}%
\patchBothAmsMathEnvironmentsForLineno{multline}%
}

\usepackage{color}
\usepackage{soul}
\usepackage{siunitx}
\usepackage[numbers,sort]{natbib}
\newcommand\norm[1]{\left\lVert#1\right\rVert}

\def\RR{ \mathbb R}

\title{Space-time multilevel Monte Carlo methods \\ and their application to cardiac electrophysiology}
\author[1]{S.~Ben~Bader}
\author[1]{P.~Benedusi}
\author[1,2]{A.~Quaglino}
\author[1]{P.~Zulian}
\author[1]{R.~Krause}
\affil[1]{Center for Computational Medicine in Cardiology,\par
Institute of Computational Science, \par
Universit\`a della Svizzera italiana,
Lugano, Switzerland}
\affil[2]{NNAISENSE SA, Lugano, Switzerland}

\date{\small\sffamily Last update: \today}

\begin{document}

\maketitle

\begin{abstract}
We present a novel approach aimed at high-performance uncertainty quantification for time-dependent problems governed by partial differential equations. In particular, we consider input uncertainties described by a Karhunen-Lo\`{e}ve expansion and compute statistics of high-dimensional quantities-of-interest, such as the cardiac activation potential. Our methodology relies on a close integration of multilevel Monte Carlo methods, parallel iterative solvers, and a space-time discretization. This combination allows for space-time adaptivity, time-changing domains, and to take advantage of past samples to initialize the space-time solution. The resulting sequence of problems is distributed using a multilevel parallelization strategy, allocating batches of samples having different sizes to a different number of processors. We assess the performance of the proposed framework by showing in detail its application to the solution of nonlinear equations arising from cardiac electrophysiology. Specifically, we study the effect of spatially-correlated perturbations of the heart fibers' conductivities on the mean and variance of the resulting activation map. As shown by the experiments, the theoretical rates of convergence of multilevel Monte Carlo are achieved. Moreover, the total computational work for a prescribed accuracy is reduced by an order of magnitude with respect to standard Monte Carlo methods.
\end{abstract}
\newpage




\section{Introduction}
Many phenomena in science, engineering, and medicine can be modeled as initial boundary value problems for an unknown function. Their numerical solution is a computationally demanding task. In addition to that, the input data or the parameters of these models might not be known exactly, i.e., they are subject to a possibly large uncertainty. In order to guarantee that the outcome of simulations can be relied upon, it is therefore imperative to quantify the effects of such input uncertainties on the output quantities-of-interest. This process is called forward propagation and is part of the field of Uncertainty Quantification (UQ). This can be achieved via, e.g sampling methods by solving, numerous times, a computationally expensive problem. This setting requires that one should not only rely on huge computational resources (e.g. parallel implementation on supercomputers), but also on mathematical state-of-the-art techniques for reducing the computational load. 

In the case of cardiac electrophysiology, it has been well established that the heart muscle-fibers play a major role in the electrical conductivity. From CT and MRI imaging, the fibers orientations as well as the conductivities they carry are extremely hard to determine precisely. Moreover, these can vary from one patient to another, leading to huge uncertainties if standard orientations or conductivities are used in patient-specific simulations. In this work, we focus only on fiber conductivities and model it as a spatially-correlated random perturbation applied to the diffusion field. In this context, a single patient-tailored simulation can take several hours on a large cluster, for example when the monodomain model is employed~\cite{niederer2011}. It follows that, e.g., UQ for such models is currently unfeasible, or extremely inefficient, with plain Monte Carlo methods.

A standard approach to overcome these difficulties is to use model reduction strategies, such as reduced or surrogate models~\cite{gian, corrado2015identification}. These approximations are build upon the observation that in many applications it is clearly possible to distinguish between \emph{offline} and \emph{online} phases of the workflow, where the former is typically very expensive and encompasses everything that can be precomputed in advance, e.g., training a surrogate model before any patient-specific data becomes available, while the latter is very cheap and consists only of model evaluations. Unfortunately, this approach has two significant drawbacks. On the one hand, complex error estimates must be provided to ensure that the approximation error is within acceptable bounds. On the other hand, the offline training should be performed only once, which is unrealistic in cases where, e.g., the geometry is not known a priori. Alternative to that, physics-based reduction strategies which bypass the precomputation are also possible. In the case of electrophysiology, the activation time can be computed via the eikonal equation, which provides a physiologically-meaningful solution~\cite{pullan2002}, strongly linked to the bidomain equation~\cite{colli93}. 

A more recent idea is to allow reduced models to be inaccurate, in the sense of not providing an approximation within certain error bounds.  Accuracy is replaced by correlation~\cite{peherstorfer2016survey} or even only statistical dependence~\cite{koutsourelakis2009accurate} with the high-fidelity model. This approach has a twofold advantage. On the one hand, it is the statistical dependence, rather than the error bounds of coarse models based on a coarse meshes, that is crucial to ensure that propagating uncertainties via the low-fidelity models provides useful information on the statistics of the high-fidelity quantity-of-interest. On the other hand, low-fidelity models are not restricted to low-resolution geometries, therefore solving them can be several orders of magnitude faster than a coarse model. However, some computational effort ($\sim$100 runs of the high-fidelity model) is needed for estimating the statistical dependence between the models. Moreover, there is no established approach for treating vector-valued quantities-of-interest~\cite{peherstorfer2015optimal}, which may require a significantly different number of samples of each model for each vector component. An approach based on minimizing a norm of the error was recently proposed in \cite{quaglino2019high}.

In the specific context of electrophysiology, the use of Bayesian multifidelity methods has been advocated in~\cite{quaglino2018fast}, where a Bayesian regression mapping the low- and high-fidelity outputs is created. The Bayesian nature of this approach error automatically augments the estimate of the solution statistics with full probability distributions and credible intervals, obtained independently of the degree of statistical dependence between models and the number of samples employed. This makes it a very suitable approach for scenarios where resources are scarce, such as 
clinical applications. However, this approach is limited to scalar or low-dimensional quantities-of-interest. Moreover, it does not yield asymptotic convergence to point estimates as the number of samples tends to infinity, so it is unsuitable for scenarios where accurate estimates are desired.

All of the above methodologies are characterized by a clear segregation among sampling, models, discretizations, and solution methods. Here, we take the opposite route and intertwine the different components. The central point of our approach is to consider the multilevel Monte Carlo method~\cite{gilesmultilevel}, which extends the idea of control variables~\cite{fishman_monte_2003}. The key idea is to perform most of the simulations on a sequence of low-resolution models. Similarly to the case of multifidelity, the direct use of the highest-resolution level guarantees convergence, while a significant portion of the computational load is offset to the low-resolution hierarchy. However, in this case, no offline preprocessing is needed to find the correlation across models and vector-valued outputs are naturally handled by the framework. In order to allow for arbitrary coarsening in space and time, we further rely on a space-time discretization of the monodomain equation. By doing so, it also becomes possible to use the already computed samples for the initialization of Newton's method, which solves the space-time equations, thereby significantly reducing the amount of iterations required for convergence.

The outline of the paper is as follows: Section 2 introduces the electrophysiology equations and the uncertainty related to the fiber conductivities, Section 3 details the proposed method using the heat equation as a model problem, and Section 4 presents the results from the numerical experiments.

%
%

\section{Modelling the uncertainty in the fiber conductivities}
\subsection{The deterministic monodomain equation}
The heart assumes a pumping function that is the result of a very complex contraction and relaxation cycle occurring in the cardiac cells. This process is controlled by a non-trivial pattern of electrical activation which is at the core of the heart function. Indeed, several heart diseases are closely related to disturbances of the electrical activity. As a consequence, understanding and modelling this activity is important for a better clinical diagnosis and treatment of patients.  

 The propagation of the electrical potential inside the cardiac muscle is initiated in the sinoatrial node, that is on top of the left and right atriums. It generates a stimulus that gives rise to a travelling wave through the heart. Its cells have the ability to respond actively to this electrical stimulus through voltage-gated ion channels.

 Starting from the first models for  ionic channels \cite{hodgkin1952quantitative}, a large family of models has been developed by now, describing the electrical activity at the subcellular level. Combined with diffusion in space, the so-called monodomain equation has been derived \cite{hurtado2014gradient,miller1978simulation} for the electrical propagation in the cardiac tissue. The monodomain equation is written as a non-stationary reaction-diffusion equation

\begin{alignat}{2} \label{monodomain}
  \dfrac{\partial u(\textbf{x}, t) }{\partial t}  -  \nabla \cdot ( G(\textbf{x}) \nabla u(\textbf{x}, t) ) + I_\text{ion}(u(\textbf{x}, t)) & = I_\text{app} (\textbf{x} ,t), && \text{  } \forall (\textbf{x},t) \in D \times (0,T], \nonumber \\
    G (\textbf{x}) \nabla u (\textbf{x}, t) \cdot \textbf{n}(\textbf{x}) & = 0, && \text{  } \forall (\textbf{x},t) \in \partial D \times (0,T],\\ 
     u(\textbf{x},0) & =  0, && \text{  } \forall \textbf{x} \in D,  \nonumber
\end{alignat}
where $u=u(\textbf{x}, t)$ is the electrical potential, $D \subset \mathbb{R}^d$ a domain representing the heart, $T\in \mathbb{R}^+$ the end time, $I_\text{app}:D \times [0,T] \rightarrow \mathbb{R}$ an applied stimulus modeling the sinoatrial node activation, and $I_\text{ion}:\mathbb{R} \rightarrow \mathbb{R}$ an ion channel model. We rely here on the Fitz-Hugh Nagumo (FHN) model \cite{fitzhugh1961impulses}, for which we have: 

\begin{equation}
I_\text{ion}(u) := \alpha (u-u_\text{rest})(u-u_\text{th})(u-u_\text{peak}).
\label{ionic_term}
\end{equation}
The values $u_\text{rest}$,$u_\text{th}$ and $u_\text{peak}$ are characteristic potential values representing the action potential behaviour that a heart cell goes through during its activation, These are respectively the resting potential $u_\text{rest}$ (cell is unactivated), the threshold potential $u_\text{th}$ (cell is triggered) and the peak value $u_\text{peak}$ (cell is activated). Here, $\alpha$ is a non-negative scaling parameter.

 Finally, $G$ in (\ref{monodomain}) represents the conductivity tensor modeling the fibers, along which electrical potential propagation takes place, all over the heart muscle. The tensor $G$ plays a major role in our model as it may represent a source of uncertainty  with regards to the fibers orientations and conductivities it models. We here assume that $G$ is a scalar isotropic diffusion field and focus exclusively on the fiber conductivities.

\subsection{The stochastic monodomain equation} \label{stochastic_monodomain}
The mathematical and numerical instruments developed in decades of research in the electrophysiology field allow in principle for virtual therapy planning. The monodomain equation is by now an established model in cardiac electrophysiology, describing with high accuracy the electrical activity in the myocardium. Nonetheless, patient-specific simulations are still not employed as a routine tool in the treatment of patients.

 A particular reason for this can be found in the data which is acquired in clinical practice. For instance, the fiber conductivities still represent a great challenge to be determined from an imaging point of a view. One should therefore account for possible uncertainties of the diffusion field $G$ in ($\ref{monodomain}$).

\vspace{0.5cm}
 Modeling the uncertainty requires to account for an additional stochastic variable in the formulation of the monodomain equation. Let us denote this stochastic variable with $\omega \in \Omega$ where $\Omega$ is the set of all possible outcomes, which in our application would represent the set of all diffusion fields modelled by the uncertainty. We are now interested in estimating the statistics of $u(\textbf{x},t,\omega) $, i.e. $\mathbb{E} [u(\textbf{x},t)] = \int_{\Omega} u(\textbf{x},t,\omega) \, d\mathbb{P}(\omega)$, as a solution to the monodomain stochastic PDE, for which the main equation reads : for almost every $\omega \in \Omega$:

\begin{equation} \label{eq:monodomain-UQ}
    \dfrac{\partial u(\cdot,\omega)}{\partial t}  -  \nabla \cdot ( G(\cdot,\omega) \nabla u(\cdot,\omega) ) + I_{\text{ion}}(u(\cdot,\omega)) = I_{\text{app}}(\cdot) , \text{  } \text{ in } D \times (0,T],
\end{equation}
 where we omitted the space and time variables $(\textbf{x},t)$ for the sake of readability. For evaluating $\mathbb{E} [u(\textbf{x},t)]$ given $(\ref{eq:monodomain-UQ})$, we rely on Monte-Carlo and Multilevel Monte-Carlo techniques as described in Section $\ref{section:methods}$. These methods imply a deterministic reformulation of $(\ref{eq:monodomain-UQ})$, and solving the monodomain equation for different independent and identically distributed samples. Naturally, parallelization plays a major role in the performance of both methods, given the important number of samples to consider.

%
%

\section{Space-time multilevel Monte Carlo} \label{section:methods} 
\subsection{Multilevel Monte Carlo}
As introduced in the previous section, we are interested in equations that are subject to uncertain or unreliable data in the aim of evaluating the information deriving from the process of simulations. This concept had been summarized under the terminology of $\emph{uncertainty quantification}$ (UQ) and is concerned about extracting robust and qualitative information despite the presence of this variability.  In the next subsection, we briefly detail the problem we intend to solve, and introduce the reader to the notation we rely on. 

\subsubsection{Problem setting and Uncertainty modelling} \label{subsection:problem-setting}
From a mathematical point of view, the underlying task resides in computing the solution of a PDE with coefficients depending on a random input belonging to a stochastic space. For the problem $\eqref{eq:monodomain-UQ}$ to be well-defined, we need to introduce the following setting.

Let $(\Omega,\Sigma,\mathbb{P})$ be a complete and separable probability space, with $\Sigma \subset 2^{\Omega}$ a $\sigma$-field and $\mathbb{P}$ a probability measure. We consider a bounded, Lipschitz domain $D \subset \RR^{d}$ with $d=1,2,3$.  We also require the Bochner space 

\[ L^2(\Omega,\Sigma,\mathbb{P};B) := \big\{  v: \Omega \rightarrow  B  | v \text{ strongly measurable such that } \norm{v}_{L^2(\Omega;B)} < \infty       \big\} ,\]

where $B = L^2([0,T];H^1(D)) $ is itself a Bochner space ($[0,T]$ denotes the time domain) and 

\begin{equation}
     \norm{v}_{L^2(\Omega;B)} := \big(  \int_{\Omega} \norm{v(\cdot,\omega)}_B^2 d\mathbb{P}(\omega)    \big)^{1/2} . 
\end{equation}
The norm on the Bochner space $B = L^2([0,T];H^1(D))$ is in turn defined as 

\begin{equation} \label{eq::norm-Bochner}
     \norm{v(\cdot,\omega)}_B := \big(  \int_{[0,T]} \int_{D} v(x,t,\omega)^2 dx dt    \big)^{1/2} . 
\end{equation}
For the sake of brevity, we will always refer to $ L^2(\Omega,\Sigma,\mathbb{P};B)$ with  $L^2(\Omega;B) $. Having everything at hand, we introduce a simplified equation model of $\eqref{eq:monodomain-UQ}$, given by

 find $\tilde{u} \in  L^2(\Omega; L^2([0,T];H^1(D))) $ such that for almost every $\omega \in \Omega,$
\begin{equation}  \label{eq:general-model}   
    \partial_t \tilde{u} (X,\omega) - \text{div} (G(X,\omega) \nabla \tilde{u} (X,\omega) ) = f(X) \text{   in } D \times [0,T],
\end{equation}
 where $X=(\textbf{x},t)$ and $G(X,\omega)$ is the diffusion field subject to the uncertainty $\omega \in \Omega$, that can be interpreted as a particular diffusion field from all the possible ones we allow for the problem for. Notice that if the forcing function $f$ was depending on $\tilde{u}(X)$ and written as $f(X) = I_{\text{app}}(X) - I_{\text{ion}}(\tilde{u}(X,\omega))$, we recover the stochastic monodomain equation  $\eqref{eq:monodomain-UQ}$. We will however later on for error estimates (cf. subsection $\ref{subsection::error-estimates}$) rely on a classical forcing function $f(X)$, independent from $\tilde{u}$, treating therefore the convergence rates for the quadrature methods on the stochastic heat equation. The PDE further encodes suitable boundary and initial conditions. We are interested in computing the statistics of the solution $u$ to the stochastic PDE $(\ref{eq:general-model})$, i.e. $\mathbb{E}[\tilde{u}](X) = \int_{\Omega} \tilde{u}(X,\omega) \, d\mathbb{P}(\omega) \in  L^2([0,T];H^1(D)).$ 

 More specifically, every $\omega \in \Omega$ expresses a different diffusion field $G(X,\omega) $. In our particular case, it is more convenient to think of these different diffusion fields as a superposition of a mean diffusion field $G_0(X)$ and a random perturbation all over the domain $D$ (random field). 

\begin{figure}[H] 
	            \centering
	            \includegraphics[width=0.75\textwidth]{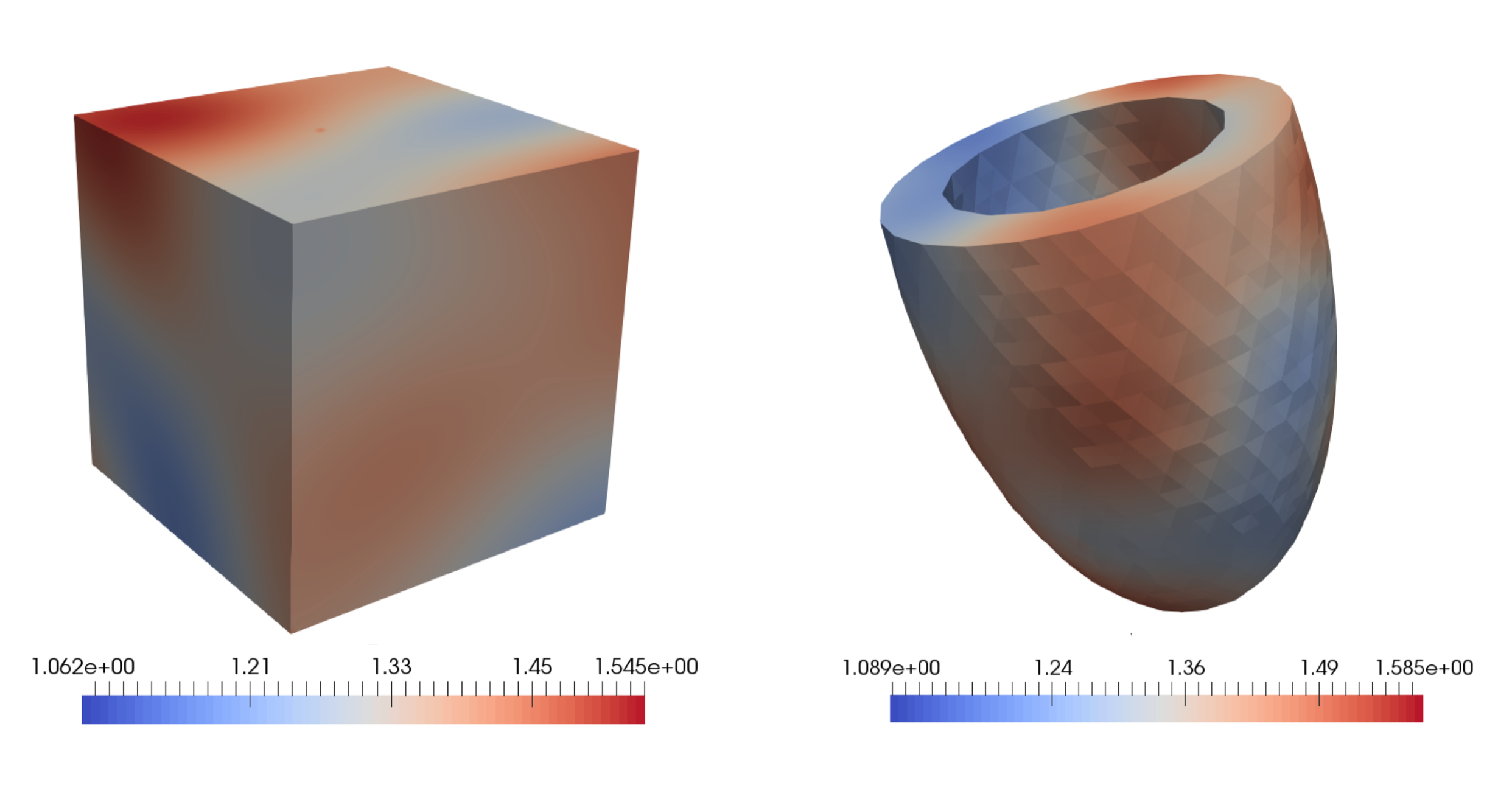}
                 \caption{Example of random fields for cube and ventricle geometries.}
                 \label{RandomField}
                 \end{figure}

 We furthermore assume the diffusion uncertainties to be spatially correlated, since neighboring regions are more likely to have the same conductivity in a healthy heart. We therefore talk about spatially correlated random fields \cite{harbrecht2015efficient}. Examples of random fields of cube and idealized ventricle geometries can be found in Figure $\ref{RandomField}$. 

The following are obtained by expressing the stochastic diffusion coefficient $G:D \times \Omega \rightarrow \mathbb{R}$ as a truncated Karhunen-Lo\`{e}ve (KL) expansion~\cite{quaglino2018fast}
\begin{equation} \label{eq:KL-expansion}
    G(\textbf{x};\omega) = G_0(\textbf{x}) + s \sum_{k=1}^{M} \sqrt{\lambda_k} \phi_k(\textbf{x}) \psi_k (\omega) \, , 
\end{equation}
where $\lambda_k$ and $\phi_k(\textbf{x})$ are respectively eigenvalues and eigenvectors of the covariance matrix induced by a spatial correlation kernel $k$. Regarding the KL truncation, the stochastic dimension $M$ is obtained by low rank approximating the covariance matrix with pivoted Cholesky decomposition~\cite{harbrecht2012,harbrecht2015efficient}. The $\left\{ \psi_k(\omega) \right\}_{k=1}^M$ are stochastically uncorrelated random variables of the joint density function $\rho (\textbf{y}) = \prod_{i=1}^M \rho_k(y_k).$ We use here a uniform distribution, i.e $\psi_i \sim \mathcal{U} (-1,1) .$ We are therefore able to identify the stochastic space $\Omega$ with its image $[-1,1]^M$ using the mapping 
\begin{alignat}{1} 
 \psi : \Omega & \rightarrow [-1,1]^M ,  \\
  \omega & \mapsto (\psi_1 (\omega) , \cdots, \psi_M (\omega) ) .  
\end{alignat}
Likewise by inserting a realization $\textbf{y} =  (y_1 , \cdots, y_M ) =  (\psi_1 (\omega) , \cdots, \psi_M (\omega) )$ , we could reformulate the truncated KL expansion in a deterministic form for $G:D \times [-1,1]^M \rightarrow \mathbb{R}$ as 
\begin{equation} \label{eq:KL-expansion-deterministic}
    G(\textbf{x},\textbf{y}) = G_0(\textbf{x}) + s \sum_{k=1}^{M} \sqrt{\lambda_k} \phi_k(\textbf{x}) y_k. 
\end{equation}
Finally, $s>0$ in $(\ref{eq:KL-expansion-deterministic})$ is chosen such that for all $x \in D$
\[ s \norm{ \sum_{i=1}^M \sqrt{\lambda_i } \phi_k(\textbf{x}) y_k }_{L^{\infty} ([-1,1]^M)} \leq \delta \, G_0(\textbf{x}), \]
for $\delta = 0.5$, in order to enforce the diffusion coefficient to be uniformly elliptic and bounded, i.e for all $\textbf{x} \in D$ and for all $\textbf{y} \in [-1,1]^M$
\[  0 < G_{\min} (\textbf{x}) \leq G(\textbf{x},\textbf{y}) \leq G_{\max} (x) < \infty . \]

 The arguments above allows also for rewriting the stochastic equation $(\ref{eq:general-model})$ in a parametric form: find $u \in L^2([-1,1]^M; L^2([0,T];H^1(D)) )$

\begin{equation} \label{eq:stochastic-heat}
\partial_t u (X,\textbf{y}) - \text{div} (G(X,\textbf{y}) \nabla u (X,\textbf{y}) ) = f(X)   \text{   in } D \times [0,T],
\end{equation}
for all $ \textbf{y} \in [-1,1]^M.$ 

 The expectation of the solution for the stochastic heat equation $(\ref{eq:stochastic-heat})$ can be rewritten as a high dimensional integral

\begin{equation} \label{eq:expectation-deterministic}
    \mathbb{E}[u] (X)  =  \int_{[-1,1]^M} u(X,\textbf{y}) \rho (\textbf{y}) \, d\textbf{y} \approx \int_{\Omega} \tilde{u}(X,\omega) \, d\mathbb{P}(\omega) = \mathbb{E}[\tilde{u}] (X) .
\end{equation}

\subsubsection{Methods and error estimates} \label{subsection::error-estimates}
 In order to numerically approximate the integral $(\ref{eq:expectation-deterministic})$, one should rely on a quadrature formula of the type
\begin{equation} \label{eq:stochastic-quadrature}
    \mathcal{Q} u (\cdot) = \sum_{i=1}^{N_l} w_i u(\cdot,\xi_i) \rho(\xi_i) \, . 
\end{equation}
 Approximating a stochastic integral likewise can further be understood as a sampling method, where the samples represent the quadrature points. A well known sampling method is the Monte-Carlo estimator. It is characterized by a fairly easy implementation and is straightforward to use, but it has a low convergence rate and hence requires a large number of samples, i.e. simulations. In this work, we consider a multilevel approach as an alternative to standard Monte-Carlo.

The Monte-Carlo (MC) estimator represents a direct method for calculating the solution to the parametric problem $(\ref{eq:expectation-deterministic}).$ Its quadrature formula can be expressed as 
\begin{equation} \label{eq:MC-estimator}
    \mathcal{Q}_{\mathcal{MC}} u (\cdot) = N_l^{-1} \sum_{i=1}^{N_l} u^i(\cdot),
\end{equation}
 i.e. the mean over solution samples $u^i$, $i=1,..,N_l$ corresponding to $N_l$ independent and identically distributed realization. 
 
 The definition of the MC estimator $\eqref{eq:MC-estimator}$ would be completely auto-sufficient if a solution $u^i$ to the sample $i$ could be computed exactly. However, in the context the monodomain equation, an exact and analytical solution can not be provided and we rely on finite elements to numerically approximate it. Let us consider a nested sequence of shape regular and quasi-uniform tetrahedralizations $\{ \mathcal{T}_l \}_{l \geq 0}$ of the spatial domain $D$, each of mesh size $h_l \thicksim 2^{-l}.$ We define the continuous, piecewise linear finite element spaces 

$$     \mathcal{S}_l = \{ v_l \in C^0(D) : v_{l|T} \in \mathbb{P}_1(T), \forall T \in  \mathcal{T}_l \}.$$
We now introduce the notation 
\begin{equation}
    \mathcal{Q}_{\mathcal{MC},l} u (\cdot) = N_l^{-1} \sum_{i=1}^{N_l} u_l^i(\cdot),
\end{equation}
where $u^i_l$ is the finite element approximation of the solution $u^i$ for the sample $i$, lying in the space $\mathcal{S}_l$. Therefore, the solution obtained through the MC method is subject not only to the stochastic error depending on the number of samples used as quadrature points, but also on the discretization error inherent to solving the PDE within the context of finite elements.

 The stochastic error of the MC quadrature integration is bounded by~\cite{barth2011multi}
\begin{equation} \label{eq:stochastic-error}
    \norm{ \mathbb{E}[u] - \mathcal{Q}_{\mathcal{MC}} u}_{L^2 ([-1,1]^m;\mathcal{V}) } \leq N_l^{-1/2} \norm{u}_{L^2 ([-1,1]^m;\mathcal{V})},
\end{equation}
where $\mathcal{V} = L^2([0,T];H^1(D))$ with the norm defined as in $\eqref{eq::norm-Bochner}$. On the other hand, using linear finite elements and a second order method in time (e.g. Crank-Nicolson) we get the following error estimate for a $H^2$--regular solution in space~\cite{quarteroni2010numerical,french1993analysis,gong2013error,aziz1989continuous} 

\begin{equation} \label{eq:discretization-error}
     \norm{ u - u_l }_{\mathcal{V}}  \leq C ( h_l^{2} + \Delta t_l^2 ) \norm{f}_{\mathcal{V}},
\end{equation}
for some constant $C>0$ that depends on the initial condition and the number of time steps considered, but not on $h_l$ and $\Delta t_l$ which are respectively the spatial and time discretization parameters for a given level $l$. Putting $(\ref{eq:stochastic-error})$ and $(\ref{eq:discretization-error})$ together results in having the general error bound~\cite{barth2011multi}
\begin{equation} \label{eq:general-error-MC}
    \norm{ \mathbb{E}[u] - \mathcal{Q}_{\mathcal{MC},l} u }_{L^2 ([-1,1]^m;\mathcal{V}) }  \leq C  ( N_l^{-1/2} + h_l^{2} + \Delta t_l^2) \norm{f}_{L^2([-1,1]^m;\mathcal{V})}.
\end{equation}
 The setting for a MC estimation is therefore closely related to that of the following error estimate, as it underlines the relationship between the mesh size and the number of samples to consider on a given discretization level. Indeed, in order to not have an error that is neither dominated by discretization in stochastics or discretization in space and time, the number of samples to consider on a level $l$ is such that
\[  N_l^{-1/2} \thicksim \mathcal{O}(h_l^2) \implies N_l \thicksim \mathcal{O}(h_l^{-4}) = \mathcal{O}(2^{4 l }),    \]
 where we assume that $h_l \sim 2^{-l}$ and $h_l \sim \Delta t_l$. The convergence rate of the MC method can therefore be derived from $(\ref{eq:general-error-MC})$ and is given by  $2^{-2l}.$ The reader might at this level realize the limitation of such a method, as the number of samples to be taken gets tremendously large when considering discretization levels with small $h_l$ and $\Delta t_l$. In addition, each one of these samples also requires a more significant computational load which places the MC method in the range of numerically costly methods for PDEs requiring a high resolution solution.   

The Multilevel Monte Carlo (MLMC) estimator represents an alternative for the MC estimator. We would like to solve the equation for an initially discretized domain $D \times [0,T]$, which we will refer to as being the fine discretization level $L$. Let us establish a hierarchy of nested coarse discretization levels $l=0,1,..,L-1$ obtained by uniform coarsening in space and time. We define the MLMC-estimator as follows:

\begin{equation} \label{eq:MLMC-estimator}
    \mathcal{Q}_{\mathcal{MLMC},L} u (\cdot) =  \sum_{l=0}^L \mathcal{Q}_{\mathcal{MC}}^{L-l} (u_l(\cdot) - u_{l-1}(\cdot) ),
\end{equation}
where $\left\{  \mathcal{Q}_{\mathcal{MC}}^{l} \right\}_{l=0}^L$ is a sequence of MC quadratures with increasing precision of order $2^{-2l}$~\cite{harbrecht2012multilevel}. We also define $u_{-1} \equiv 0.$

 To clarify the above, the underlying idea about the MLMC-estimator is to perform the dominant number of MLMC samples on a coarse level $l=0$, and to obtain a solution for level $l=L$ by adding corrections between every different successive levels $l-1$ and $l$. The MLMC and its convergence rate have been discussed in details in~\cite{teckentrup2013further,charrier2013finite,harbrecht2012multilevel} for the case of the stochastic Poisson problem. For the stochastic equation, given $(\ref{eq:stochastic-error})$, $(\ref{eq:discretization-error})$ and the above specified  $2^{-2l}$ increasing precision for the quadrature rule, a similar result can shown:

\begin{equation} \label{eq:general-error-MLMC}
    \norm{ \mathbb{E}[u]  - \mathcal{Q}_{\mathcal{MLMC},l} u }_{L^2([-1,1]^m;\mathcal{V})}  \leq C 2^{-2L} L \norm{f}_{L^2([-1,1]^m;\mathcal{V})},
\end{equation}
 for some constant $C>0$. The MLMC has therefore the same convergence rate than MC up to a logarithmic number, but is designed to be less computationally expensive.

\subsection{Space-time discretization for the linear problem}\label{disc}
The heat equation \eqref{eq:general-model} can be numerically solved by means of finite elements (FE) and finite differences (FD), which are well established for such problems. A classical approach in that sense is to perform a FE discretization in space and FD in time, i.e using a sequential time-stepping method. The parallel scalability of this approach is eventually limited by the time integration process. In fact, when the parallelization in space saturates, sequential time integration becomes the natural bottleneck for the scalability of the solution process. This can be overcome by parallel-in-time methods. Those were developed in the last decades~\cite{nievergelt1964parallel} to overcome this bottleneck and enhance parallelism in both space and time. 
In particular, we employ a monolithic approach in space and time, where we assemble a large space-time system that is solved in parallel \cite{mcdonald2016simple}. For a comprehensive review of parallel-in-time methods, see \cite{gander201550}.

We use Lagrangian FE of first order for the discretization in the space variable. Assuming that the solution $u(\textbf{x} , t)$ is sufficiently regular on $D$, we derive the weak formulation from \eqref{eq:general-model} (in which we neglect the stochastic variable $\omega$) using test functions $v \in H^1_0(D) $: 

\begin{align*}
\int_D \dfrac{\partial u(\textbf{x} , t)}{\partial t} v(\textbf{x}) \, d\textbf{x}
+\int_D (G(\textbf{x}) \nabla u(\textbf{x},t)) \nabla v(\textbf{x}) \, d\textbf{x}= \int_D f(\textbf{x} , t) v(\textbf{x})  \, d\textbf{x}.
\end{align*}
 Then, we define the semi-discrete problem:

\begin{equation}
  M_h  \dfrac{\partial \textbf{u}(t)}{\partial t} + K_h\textbf{u}(t) = \textbf{f}(t).
  \label{semi-disc}
\end{equation}
 Here, $M_h \in \mathbb{R}^{n \times n}$ and $K_h \in \mathbb{R}^{n \times n}$ are the standard mass and stiffness matrices obtained using $n$ linear nodal basis functions $\{\psi_i\}_{i=1}^{i=n} \subset \mathbb{P}_1$, i.e.

\begin{equation}
  M_h: = \left[ \int_D \psi_i(\textbf{x}) \psi_j(\textbf{x})  d\textbf{x} \right]_{i,j=1}^n, \quad
  K_h := \left[ \int_D (G(\textbf{x}) \nabla \psi_i(\textbf{x})) \nabla \psi_j(\textbf{x})  d\textbf{x} \right]_{i,j=1}^n,
\end{equation}

\begin{equation}
  \textbf{f}(t) := \left[ \int_D f(\textbf{x} , t) \psi_i(\textbf{x}) \right]_{i=1}^n,
\end{equation}
 arising from the approximation:

\begin{align*}
u(\textbf{x} , t) \simeq \sum_{i=1}^n u_i(t)\psi_i(\textbf{x}), \quad \text{with} \quad \textbf{u}(t) = [u_1(t), u_2(t),\cdots,u_n(t)]^T.
\end{align*}
 Consider a uniform partition of the time axis $[0,T]$ in $m$ intervals, such that $\Delta t  = T/m$ and $t_k = k\Delta t$, with $k=0,\cdots,m$. We apply the second order Crank–Nicolson method for the time discretization of \eqref{semi-disc} and obtain, for  $k=0,\cdots,m-1$

\begin{equation}
 \left( M_h + \frac{\Delta t}{2}K_h \right) \textbf{u}_{k+1} + \left( - M_h + \frac{\Delta t}{2}K_h \right)\textbf{u}_{k} = \textbf{f}_k \quad \text{and} \quad \textbf{u}_{k}:= \textbf{u}(t_k),
 \label{st-system}
\end{equation}

\begin{equation}
  \text{with} \quad  \textbf{f}_k:= \frac{\Delta t}{2} \left(  \textbf{f}(t_{k+1}) + \textbf{f}(t_k)  \right).
\end{equation}
 If we define $A_{t,h} :=  M_h + \frac{\Delta t}{2}K_h $ and $B_{t,h} := - M_h + \frac{\Delta t}{2}K_h$ the system of equations ($\ref{st-system}$) can be summarized in the compact form: 
\begin{equation}
\begin{tikzpicture}[baseline=(current bounding box.center)]
\matrix (m) [matrix of math nodes,nodes in empty cells,right delimiter={]},left delimiter={[} ]{
A_{t,h}  &  &   &    \\
B_{t,h}  & A_{t,h} & &  \\
 & & &     \\
  & & B_{t,h} & A_{t,h}  \\
} ;
\draw[loosely dotted] (m-2-1)-- (m-4-3);
\draw[loosely dotted] (m-2-2)-- (m-4-4);
\end{tikzpicture}\begin{pmatrix}
    \textbf{u}_1\\\textbf{u}_2\\\vdots\\\textbf{u}_{m}
\end{pmatrix}=
\begin{pmatrix}
    \textbf{f}_1\\\textbf{f}_2\\\vdots\\\textbf{f}_{m}
\end{pmatrix} 
 \quad \Longleftrightarrow  \quad C \textbf{u} = \textbf{f},
\label{eq:system_C}
\end{equation}
 where $C\in \mathbb{R}^{nm \times nm}$ is a large space-time system that can be distributed and solved in parallel and 
\[ \textbf{u} := [\textbf{u}_1,\textbf{u}_2,\cdots,\textbf{u}_{m}]^T \quad \text{and} \quad \textbf{f} := [\textbf{f}_1,\textbf{f}_2,\cdots,\textbf{f}_{m}]^T.\] 

\subsection{Discretization of the nonlinear problem}
The discretization of \eqref{monodomain} is an extension of the assembly procedure described in Section~\ref{disc}.  In particular, the linear system \eqref{eq:system_C} is modified to contain the discretization of the non-linear reaction term $I_{\text{ion}}$:
\begin{equation}\label{mono_disc}
C\mathbf{u} + \mathbf{r}(\mathbf{u}) = \mathbf{f},
\end{equation}
where $\mathbf{r}(\mathbf{u})\in \mathbb R^{nm}$ is given by 
\begin{equation*}
   \mathbf{r}(\mathbf{u}):=(\Delta t I_N \otimes M_h) \mathbf{I}_{\text{ion}}(\mathbf{u}) \quad \text{with} \quad   \mathbf{I}_{\text{ion}}(\mathbf{u}):=[I_{\text{ion}}(u_1),I_{\text{ion}}(u_2),\cdots,I_{\text{ion}}(u_{nm})]^T, 
\end{equation*}
where $n$ and $m$ are respectively the space and time degrees of freedom. \\
The Jacobian $\mathbf{J}(\mathbf{u})\in\mathbb R^{nm\times nm}$ of the non-linear operator in left hand side of \eqref{mono_disc} is given by
\begin{equation*}
    \mathbf{J}(\mathbf{u})= C + (\Delta t I_m \otimes M_h)\cdot\mathrm{diag}(I_{\text{ion}}'(u_1),I_{\text{ion}}'(u_2),\cdots,I_{\text{ion}}'(u_{nm})). 
\end{equation*}

\subsection{Transfer of discrete fields}
The transfer of information between finite element spaces in our multilevel Monte Carlo method is performed by means of both $L^2$--projections and standard finite element interpolation methods. 
In the case when mesh hierarchies are generated by refinement, the elements of the refined mesh form partitions of the elements of the original coarse mesh. This allows us to employ standard finite element interpolation for transferring from the coarse space to the refined space. However, if the fine mesh is not generated by refinement we can use $L^2$--projections for the same task. For transferring discrete fields from the fine space to a coarser space we also employ the $L^2$--projection. $L^2$--projections are proven to be optimal and stable and in general superior to interpolation~\cite{hesch}. In particular, we employ a local approximation of the $L^2$--projection, which is constructed by exploiting the properties of the dual basis~\cite{wohlmuth_2000,popp_2012}. This local approximation allows us to to construct the transfer operator explicitly in such a way that it can be applied by means of a simple sparse matrix-vector multiplication.
In a parallel computing environment, where meshes are arbitrarily distributed, the assembly of the mass matrices related to $L^2$--projection is not trivial. The construction of the discrete $L^2$--projection requires us to detect and compute intersections between the elements of the coarse mesh and the elements of the finer mesh which might be stored in different memory address spaces (e.g., on a super-computing cluster). For this purpose we use the parallel tree-search algorithms and assembly routines described in~\cite{krause2016parallel}. 

In this work we extend the above to multilevel space-time discretizations. 
By exploiting the tensor-product structure of~\eqref{eq:system_C} we can simplify the implementation and benefit from better computational performance.
As before, let us assume that the spatial mesh does not change at each time-step (i.e., no adaptive mesh refinement). In this case, the tensor-product structure of the space-time grid allows us to construct space-time operators in a convenient way 
which requires the assembly of the spatial transfer operator to be performed only once.

For $l=0,...,L-1$ let $\mathbf{u}^l \in \mathbb R^{n_lm_l}$ be the coefficients of the discrete space-time solution on level $l$, where $n_l$ (resp. $m_l$) is the number of spatial nodes (resp. time steps) on $l$. With $P_{l,h} \in \mathbb{R}^{n_{l} \times n}$ we denote the spatial discrete representations of the $L^2$--projection from the fine level to a coarser level $l$, and  with $P_{l,t} \in \mathbb{R}^{m_{l} \times m}$ the temporal restriction operator, obtained by standard bisection. The space-time transfer matrix $P_l \in \mathbb{R}^{n_lm_l \times nm}$ is constructed with the following tensor product \[P_l = P_{l,t} \otimes P_{l,h},\] 
 and we write $ \mathbf{u}^{l} = P_l \mathbf{u}$. The space-time interpolation (prolongation) operator from any coarse level $l-1$ to a finer level $l$ are constructed in the same way, using the tensor-product between transfer operators.
 
\subsection{Parallel solver}
The non linear problem \eqref{mono_disc} is solved with Newton's method. For each of the arising linear problems in the form \eqref{eq:system_C}, we employ a space-time parallel PGMRES, with the preconditioner given by an ILU(0) factorization within the PETSc \cite{petsc-user-ref} framework. If multiple processors are used, the corresponding block-Jacobi preconditioner is employed, and the PGMRES is applied, in parallel, to the local blocks. The usage of the latter PGMRES is motivated by the spectral analysis of the space-time system in \eqref{eq:system_C}, carried out in~\cite{Benedusi2018}. As default by PETSc, the GMRES solver is restarted after 30 iterations. It is important to recall that the first iterate plays an important role for the convergence of Newton's method. For this reason we use the unperturbed reference solution as initial guess for each Monte-Carlo sample. This guarantees fast convergence in our setting. 

\subsection{Parallelization strategy}
The MLMC method developed above reduces undoubtedly the computational complexity that is carried by a naive MC approach. However, to fully exploit the multilevel algorithm, it is necessary to rely on a parallel environment that takes advantage of all different layers under which concurrent work can be executed. The MLMC requires computing samples at different discretization levels, requiring a different amount of resources. Therefore, we could think of initially distributing the required amount of work for every level as a first parallelization layer. On every one of these levels, multiple samples need to be computed. Those are completly independent and we can consider distributing work across samples as a second parallelization layer. The third layer, in the context of finite elements, would naturally be a parallelization across the tempo-spatial grid, using space-time parallel method.

\begin{figure}[h!]
	            \centering
	            \includegraphics[width=0.75\textwidth]{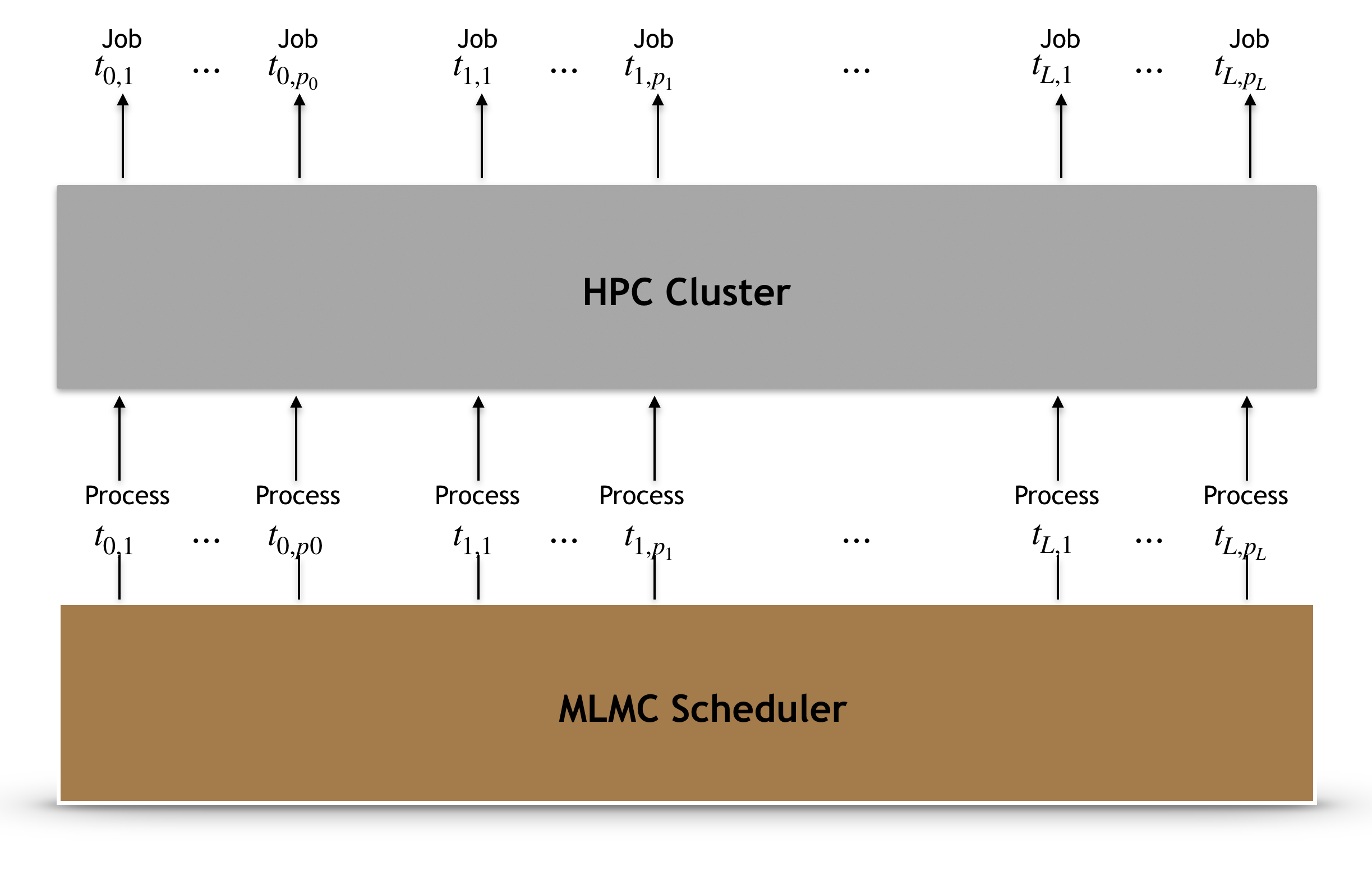}
                 \caption{MLMC tasks scheduling and job requests step. $t_{l,i}$ denotes the $i$-th task of level $l$. }
                 \label{fig::MLMC-scheduler}
                 \end{figure}

\begin{figure}[h!]
	            \centering
	            \includegraphics[width=0.75\textwidth]{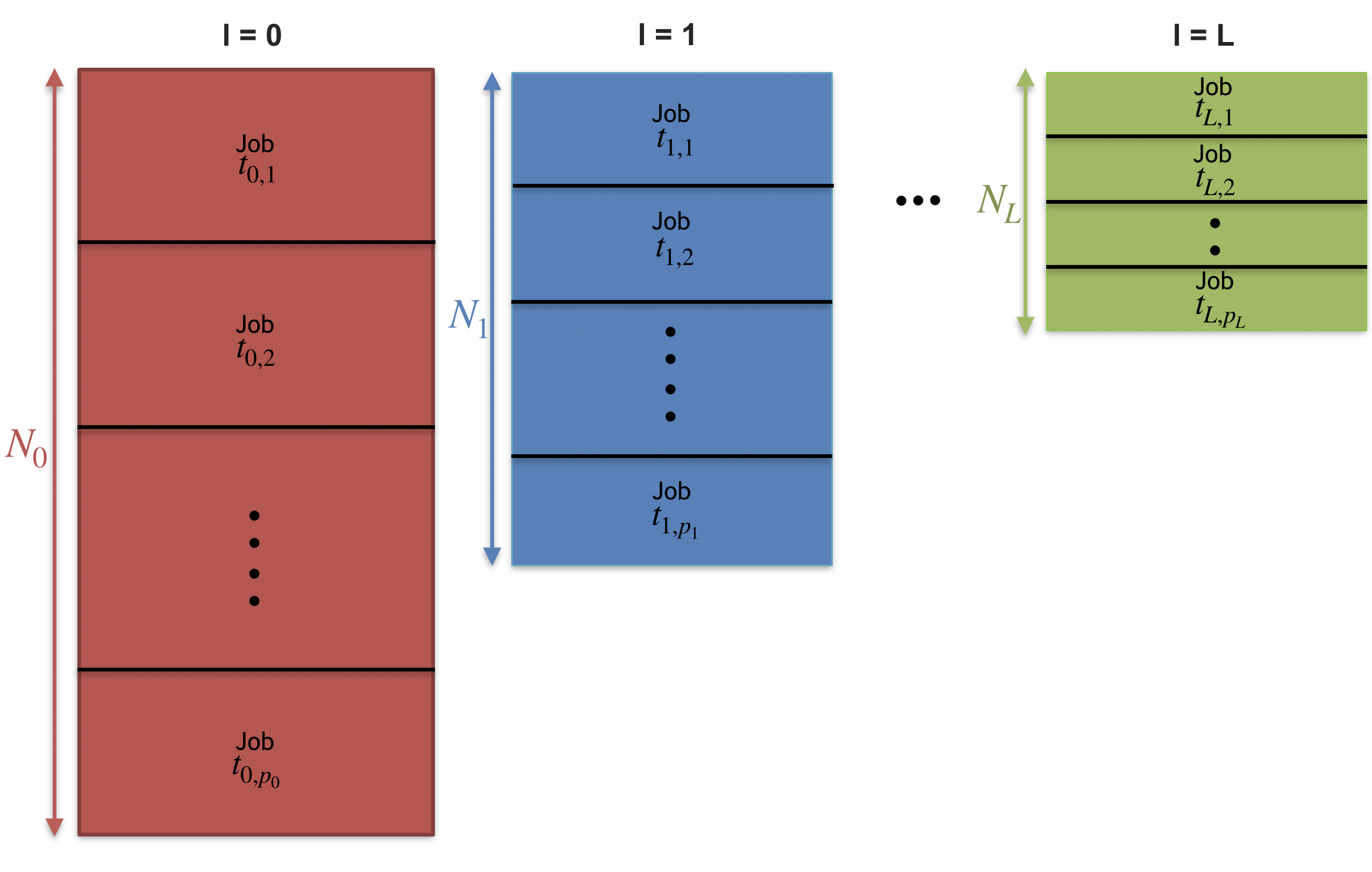}
                 \caption{The different processes executing concurrently different batches of samples.}
                 \label{fig::parallelization-strategy}
                 \end{figure}

 We therefore rely on a so-called three level parallelization strategy largely inspired by \cite{drzisga2017scheduling}. We furthermore adapt it to standard allocation procedures on current High Performance Computing (HPC) system, in which a request for a small amount of resources is more likely to be granted in a shorter amount of time. We issue consequently different job calls involving different amount of resources for every parallel task consisting of a batches of samples to solve (cf. Figure $\ref{fig::MLMC-scheduler}$). As an example, let us suppose we have an MLMC setting of $L+1$ levels. We start with initiating $p_0,\ldots,p_L$ processes carrying the tasks for every one of these levels. The samples to be computed for every level $l$ are equally distributed in batches over the $p_l$ processes created for that level. The different number of samples $N_l$ on each level and the flexibility in choosing the number of processes $p_l$ are such that the batch size may vary across the levels, as shown in Figure $\ref{fig::parallelization-strategy}$. Every sample is further solved in parallel with domain decomposition techniques, fulfilling the third parallelization layer. Details about the parallel solver we relied on are provided at the end of subsection $\ref{stochastic_monodomain}$ after introducing our application, i.e. the monodomain equation. Notice that another advantage of the following cluster scheduling approach is to avoid wasting resources when the processes have different time execution.

%
%

\section{Numerical experiments}
The numerical experiments were realized using SLOTH \cite{quaglino2017h} a UQ python library developed at the Institute of Computational Science (ICS) of Lugano. For this work, we extended it to the monodomain equation (and in general to all types of 3D+1 PDEs) by employing Utopia \cite{Zulian2017d} for the finite element formulation. The experiments have been performed on the multicore partition of the supercomputer Piz Daint of the Swiss National Supercomputing Centre.

\begin{figure}[H] 
	            \centering
	            \includegraphics[width=0.37\textwidth]{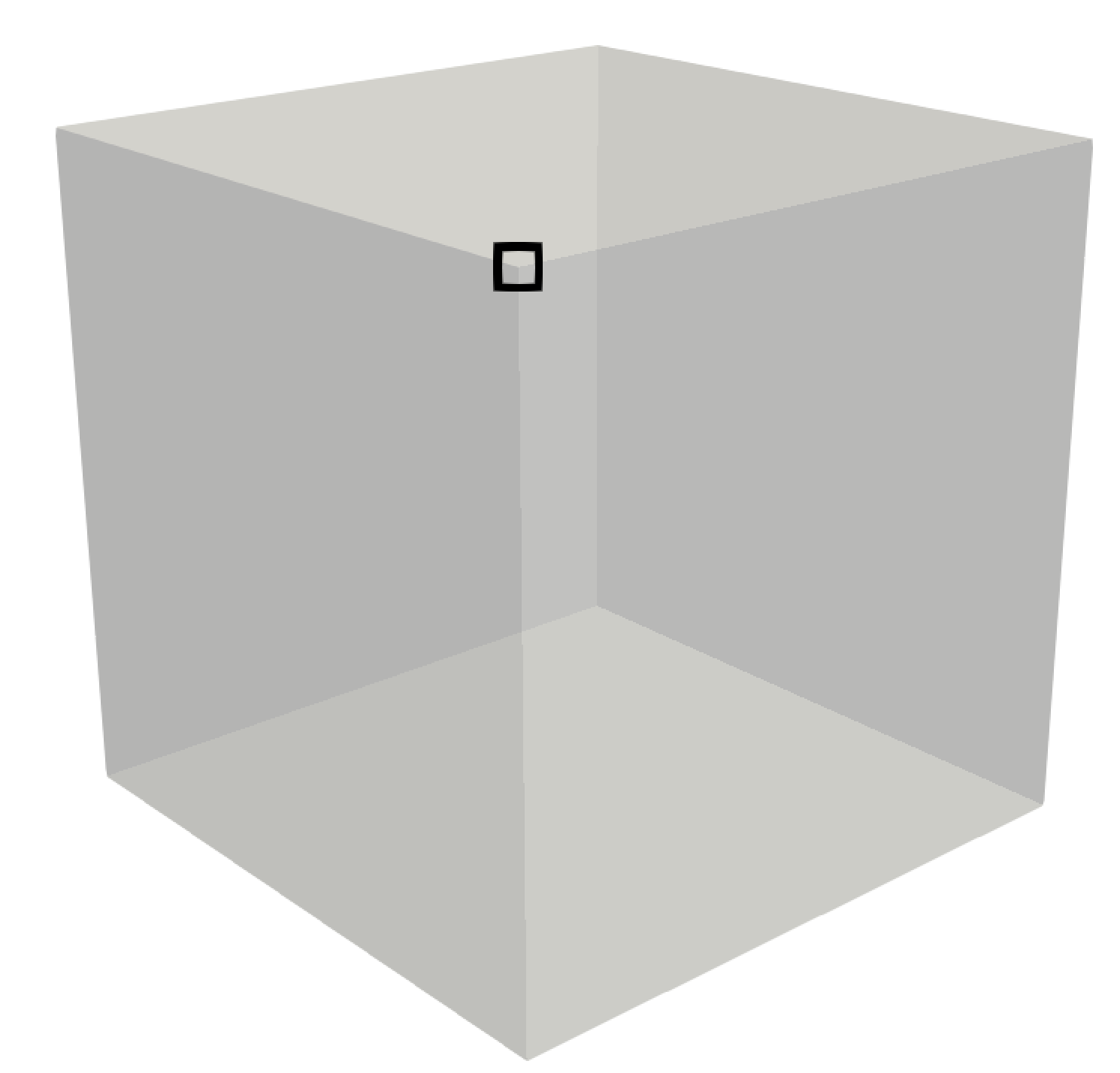}
	            \hspace{0.05\linewidth}
	            \includegraphics[width=0.37\textwidth]{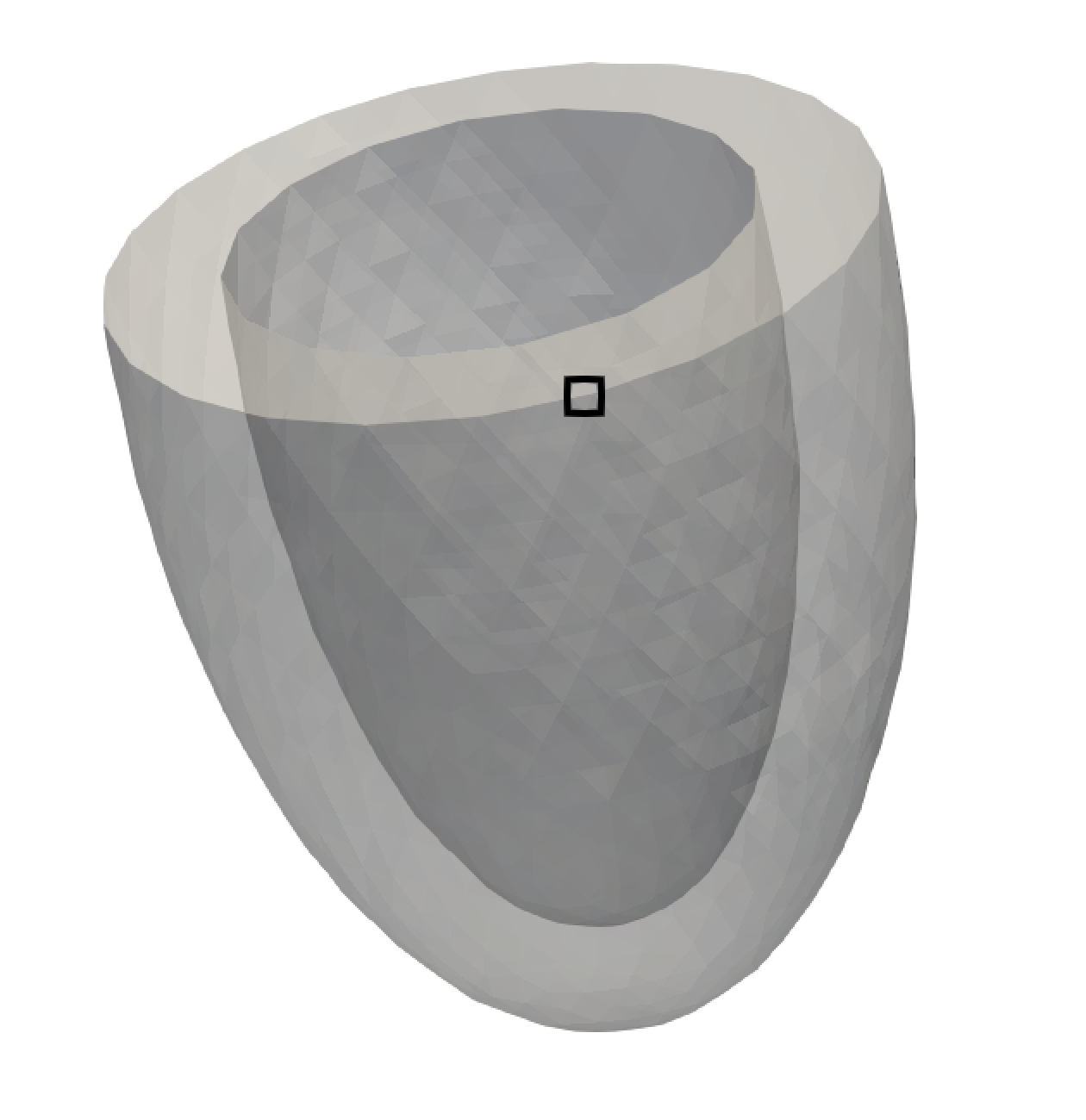}
                 \caption{Stimulus locations for the cube and the ventricle.}
                 \label{fig::stimulus-location}
                 \end{figure}

\paragraph{Parameters for the experiments} The parameters of monodomain equation $\eqref{monodomain}$ used for the numerical tests are:
    
\begin{itemize}
    \item $G_0 (\textbf{x}) = 3.325\cdot10^{-3}$cm$^2$ mV ms$^{-1}$ from $\eqref{eq:KL-expansion}$ is the mean diffusion field.
    \item $\alpha = 1.4\cdot10^{-3}$ mV$^{-2}$ms$^{-1}$, $u_\text{rest}=0$ mV, $u_\text{th}=28$ mV, and $u_\text{peak} = 115$ mV are the values for the ion channel model $I_\text{ion}(u)$ in $\eqref{ionic_term}$. 
    \item $I_{\text{app}}(\textbf{x},t) = \left( u_\text{rest} + u_\text{peak} \exp{ \left( - \dfrac{ ( \textbf{x} - \textbf{x}_0 )^2 }{ \sigma^2 } \right)} \right) \chi_{[0,t_1)}(t)$ where $t_1 = \Delta t = 0.005$ ms is the function we rely on for the applied stimulus. For the cube, $\sigma = 0.5$ cm whereas for the ventricle it was set $\sigma = 1$ cm. Here, $\textbf{x}_0$ is the location of stimulus and are shown for both geometries in Figure $\ref{fig::stimulus-location}$.

\end{itemize}

\vspace{0.2cm} 

\vspace{0.5cm}          

\begin{figure}[H] 
	            \centering
	            \includegraphics[width=0.45\textwidth]{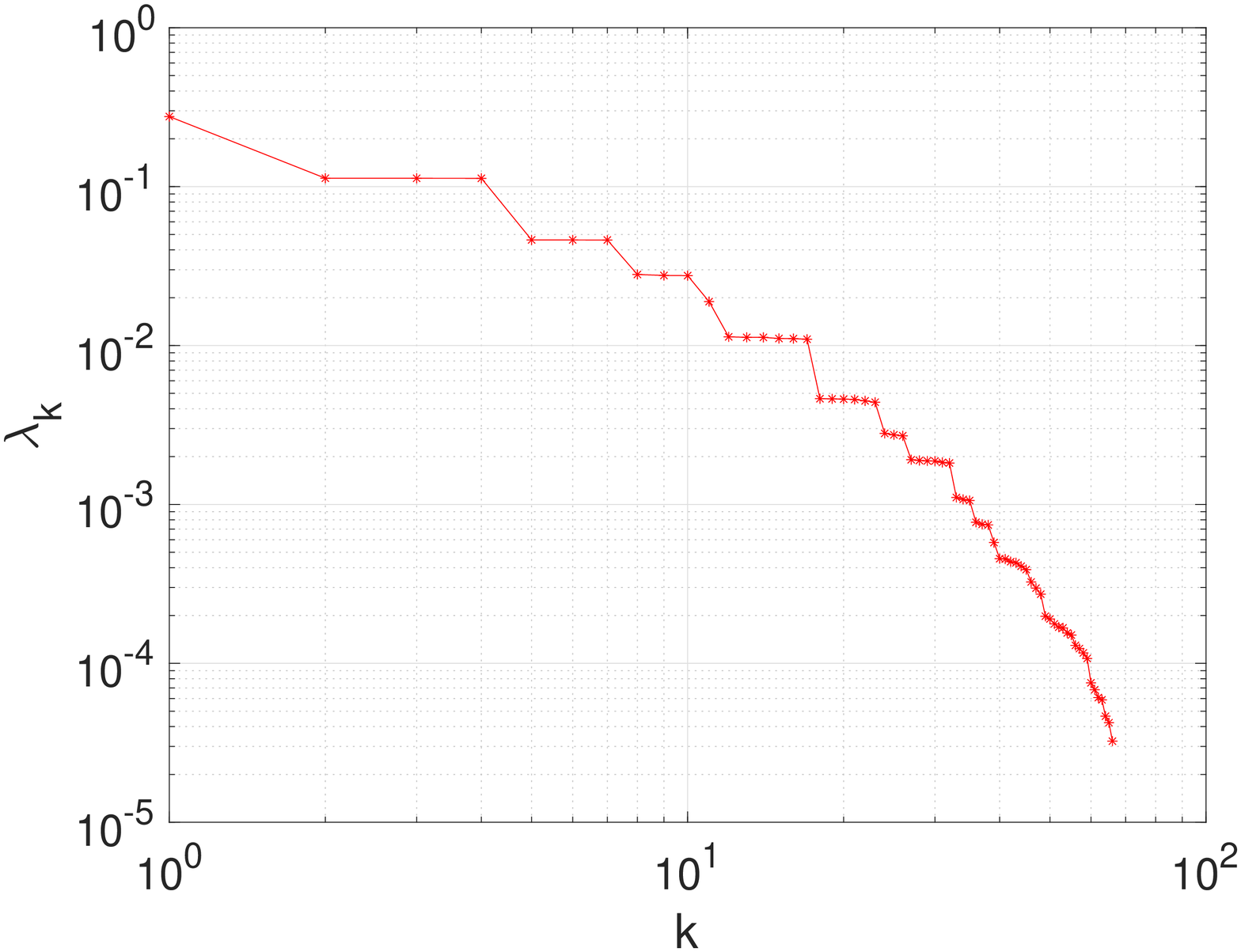}
	            \includegraphics[width=0.45\textwidth]{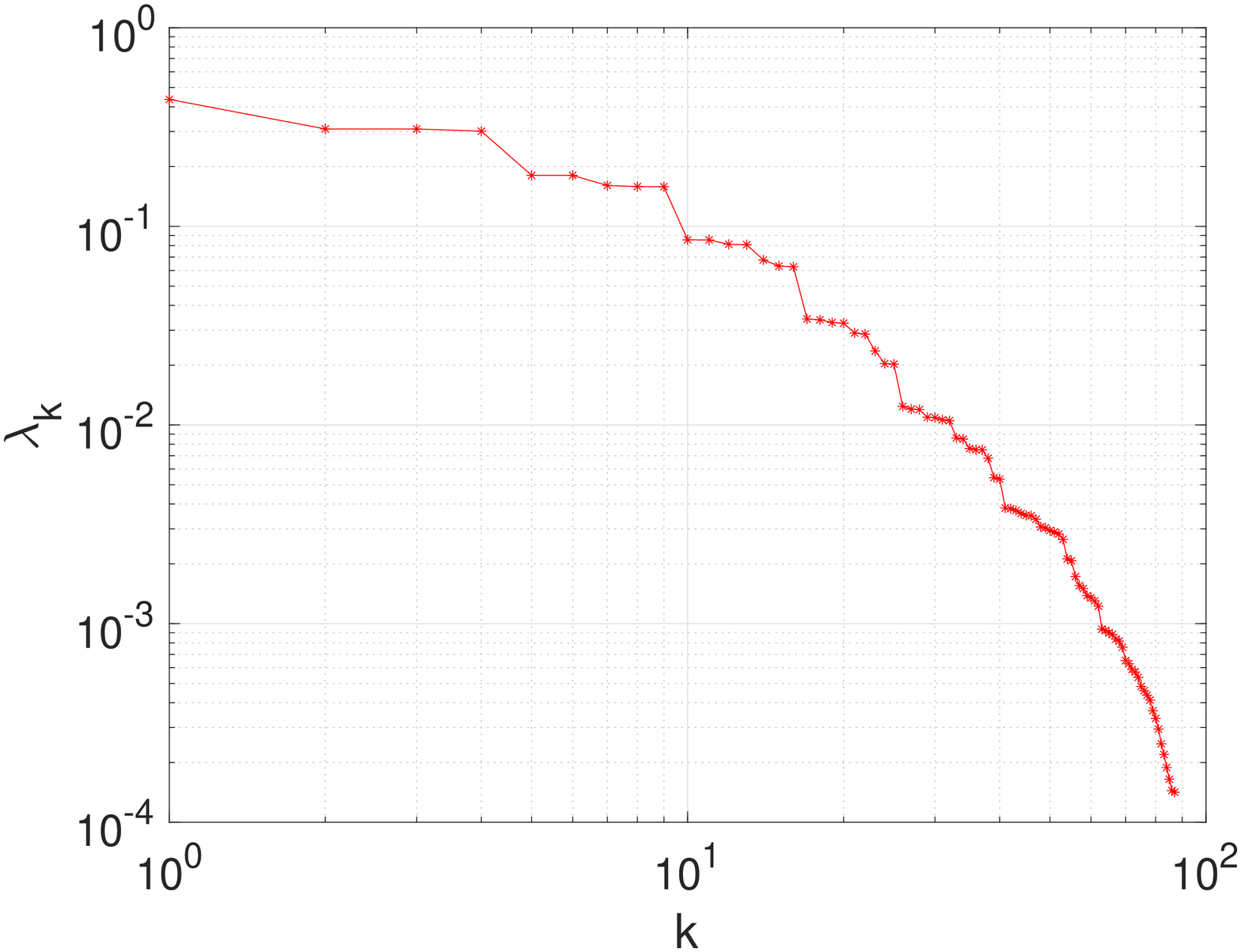}
                 \caption{KL eigenvalues decay for the cube (left) and the idealized ventricle (right).}
                 \label{fig::eigenvalues-decay}
                 \end{figure}

\paragraph{Parameters for KL expansion} We use the correlation kernel

$$ k(\textbf{x},\textbf{y}) = e^{\frac{-d(\textbf{x},\textbf{y})^2}{\sigma_{\text{KL}}}} ,$$
for $\sigma_{\text{KL}} =0.25$ for the cube and $\sigma_{\text{KL}} = 0.5$, while $d(\textbf{x},\textbf{y})$ is the euclidian distance between $\textbf{x}$ and $\textbf{y}$. The \textit{dumping factor} $s$ in $\eqref{eq:KL-expansion}$ for enforcing uniform ellipticity is set to $s=0.3 $ for both geometries. The low rank pivoted Cholesky decomposition on the covariance matrices induced by the correlation kernels above yielded the stochastic dimension $M = 66$ for the cube and $M=87$ for the idealized ventricle. We furthermore report the KL eigenvalues decay in Figure \ref{fig::eigenvalues-decay}. 

\subsection{Numerical experiments in 1D+1}
We intend to verify the convergence rate of both MC and MLMC as developed in Section $\ref{section:methods}$. We consider an interval and use a hierarchy nested meshes, obtained by uniform coarsening in space and time. In Table $\ref{table:1d-levels}$, we report all the details regarding the considered mesh hierarchy, i.e. the number of elements, the mesh size, the number of time steps and the timestep size.

\begin{figure}[h!]
	            \centering
	            \includegraphics[width=0.55\textwidth]{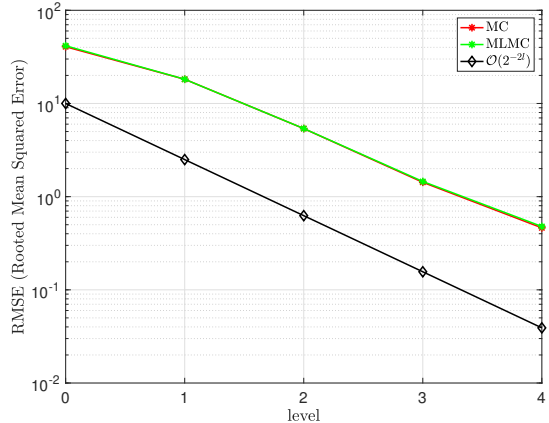}
                 \caption{Controlled convergence graphs for MC and MLMC for the 1D+1 example.}
                 \label{fig::Controlled-Convergence-1D}
                 \end{figure}

\vspace{0.8cm}
\begin{table}[h!]
\centering
\captionsetup{justification=centering}
\begin{tabular}{|c||l|l|l|l|l|l|} 
\hline $l$ & 0 &  1 & 2 & 3 & 4 & 5  \\
\hline
\hline $ n $  & 31 & 62 & 125 & 250 & 500 & 1000   \\
\hline $ h $  & 0.032 & 0.016 & 0.008 & 0.004 & 0.002 & 0.001   \\
\hline
\hline $ m $  & 4 & 8 & 16 & 32 & 64 & 128   \\
\hline $ \Delta t $  & 0.16 & 0.08 & 0.04 & 0.02 & 0.01 & 0.005   \\
\hline 
\end{tabular}

\caption{Details about the considered Mesh hierarchy. $\Delta t $ and $h$ are the time and space discretization steps, $m$ and $n$ are the number of subdivisions for time and space intervals}. 

\label{table:1d-levels}
\end{table}
\vspace{0.8cm}
We create a reference solution, denoted as $v_{ref}$, by calculating the mean over 20'000 samples on the finest level $L=5$. For the purpose of verifying the MC convergence rate, we run several MCs on each level $l$ with $N_l \sim 2^{4l},$ and compute the difference between the expectation calculated at each level, denoted as $v_l$, and the reference solution. We evaluate the rooted mean square $\text{RMSE} = \mathbb{E} [\| e_l \|^2]^{1/2}$, where $e_l = | v_l - v_{\text{ref}}|$ and 
 
 $$ \| e_l \|^2 = \int_{[0,T]} \| e_l(\cdot,t)\|^2_{L^2(D)}  \, dt = \int_{[0,T]} \int_{D}  e_l^2(x,t) \, dx  \, dt .$$
 We plot the rooted mean square error versus the level, in what we will refer to as the controlled convergence graph. We use this terminology as it indicates that we are controlling the number of samples in order to have a stochastic and discretization errors of the same order. The plot is shown in Figure $\ref{fig::Controlled-Convergence-1D}$.

 In the same plot, we show the controlled convergence for the MLMC. For each level $L$, we perform an MLMC run involving as many levels by setting the number of samples on the coarse level to $ 2^{4L}$. The number of samples for the following levels is then calculated by dividing this number with the ideal sampling ratio, which corresponds in our case to $16$. This number can be deduced by enforcing that $h_l^2 N_l^{-1/2} = h^2_{l+1} N_{l+1}^{-1/2} $ which represents the error contribution from levels $l$ and $l+1.$

  \subsection{Numerical experiments in 3D+1}
  
   We intend in the next subsections to estimate and verify the convergence behaviour for the MC and the MLMC estimators, as done for the 1D+1 case above. Furthermore, a comparison in terms of work to accuracy for both methods will be conducted.
  
   We rely on the cube and the ventricle geometries, introduced in Section $\ref{section:methods}$ (cf. Figure $\ref{RandomField}$). As opposed to a classical MLMC approach, where one would refine an initially coarse mesh in order to reduce the variance, we here consider the realistic setting of an application for which only one mesh of the given geometry is provided. This mesh is provided at the fine discretization level, therefore the goal for the MLMC is to reduce the work load by considering coarser meshes. We rely on a nested mesh hierarchy of 3 and 4 levels respectively for the ventricle and the cube geometries. The number of space-time degrees of freedom (Dof's), the space and time discretization steps of all the considered levels for the cube and the ventricle are respectively reported in Tables $\ref{table:cube-levels}$ and $\ref{table:ventricle-levels}$.
  
 \begin{table}[h!]
\centering
\begin{tabular}{|c||l|l|l|l|} 
\hline $l$ & 0 &  1 & 2 & 3  \\
\hline
\hline Dof's  &  5400 & 65'219 & 898'317 & 13'301'753   \\
\hline 
\hline 
$h$ & 0.2 & 0.1 & 0.05 & 0.025   \\
\hline
\hline
$\Delta t$ & 0.04 & 0.02 & 0.01 & 0.005   \\
\hline       
\end{tabular}
\caption{Details about the mesh hierarchy for the cube geometry.}
\label{table:cube-levels}
\end{table}

\begin{table}[h!]
\centering
\begin{tabular}{|c||l|l|l|} 
\hline $l$ & 0 &  1 & 2   \\
\hline
\hline Dof's  & 154'546 & 2'120'420 & 31'184'747   \\
\hline 
\hline 
$h$  & 0.1 & 0.05 & 0.025   \\
\hline
\hline
$\Delta t$  & 0.02 & 0.01 & 0.005   \\
\hline

\end{tabular}
\caption{Details about the mesh hierarchy for the ventricle geometry.}
\label{table:ventricle-levels}
\end{table}
  
  \subsubsection{Monte-Carlo in 3D+1}
\begin{figure}[H] 
	            \centering
	            \includegraphics[width=0.4\textwidth]{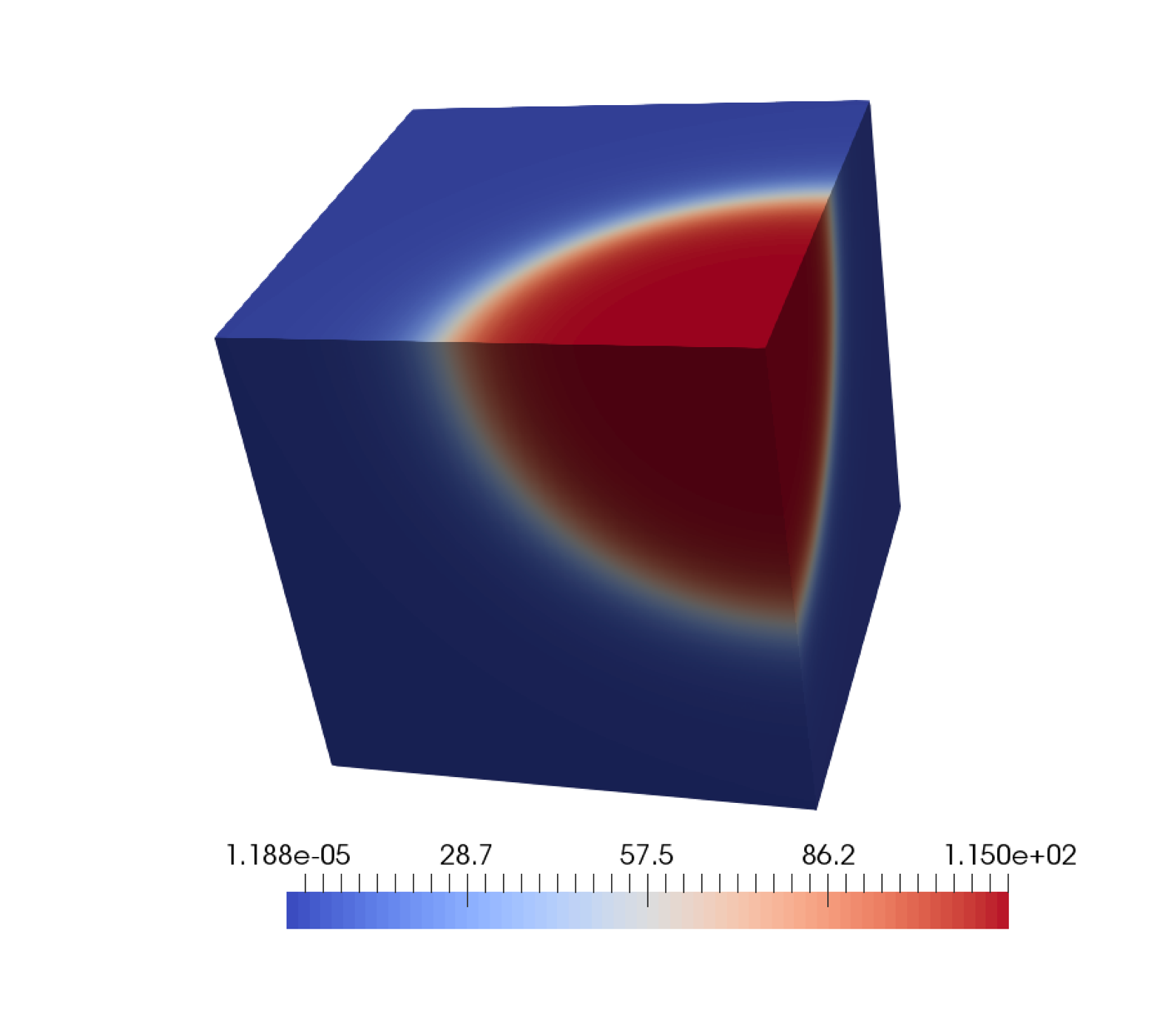}
	            \includegraphics[width=0.4\textwidth]{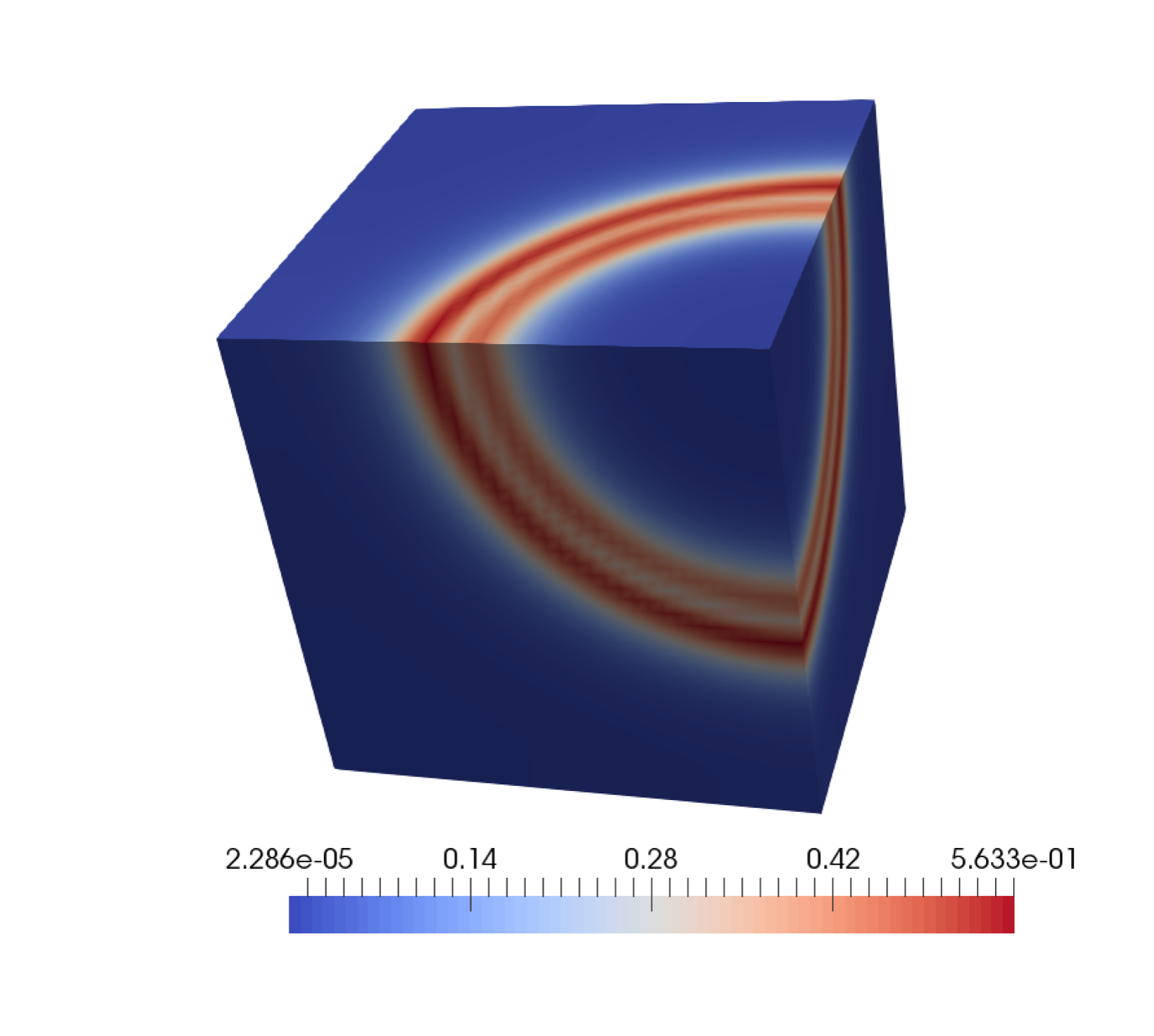}
                 \caption{Mean (left) and variance (right) at the final time state for the cube.}
                 \label{fig::MC-cube}
                 \end{figure}

\begin{figure}[H] 
	            \centering
	            \includegraphics[width=0.4\textwidth]{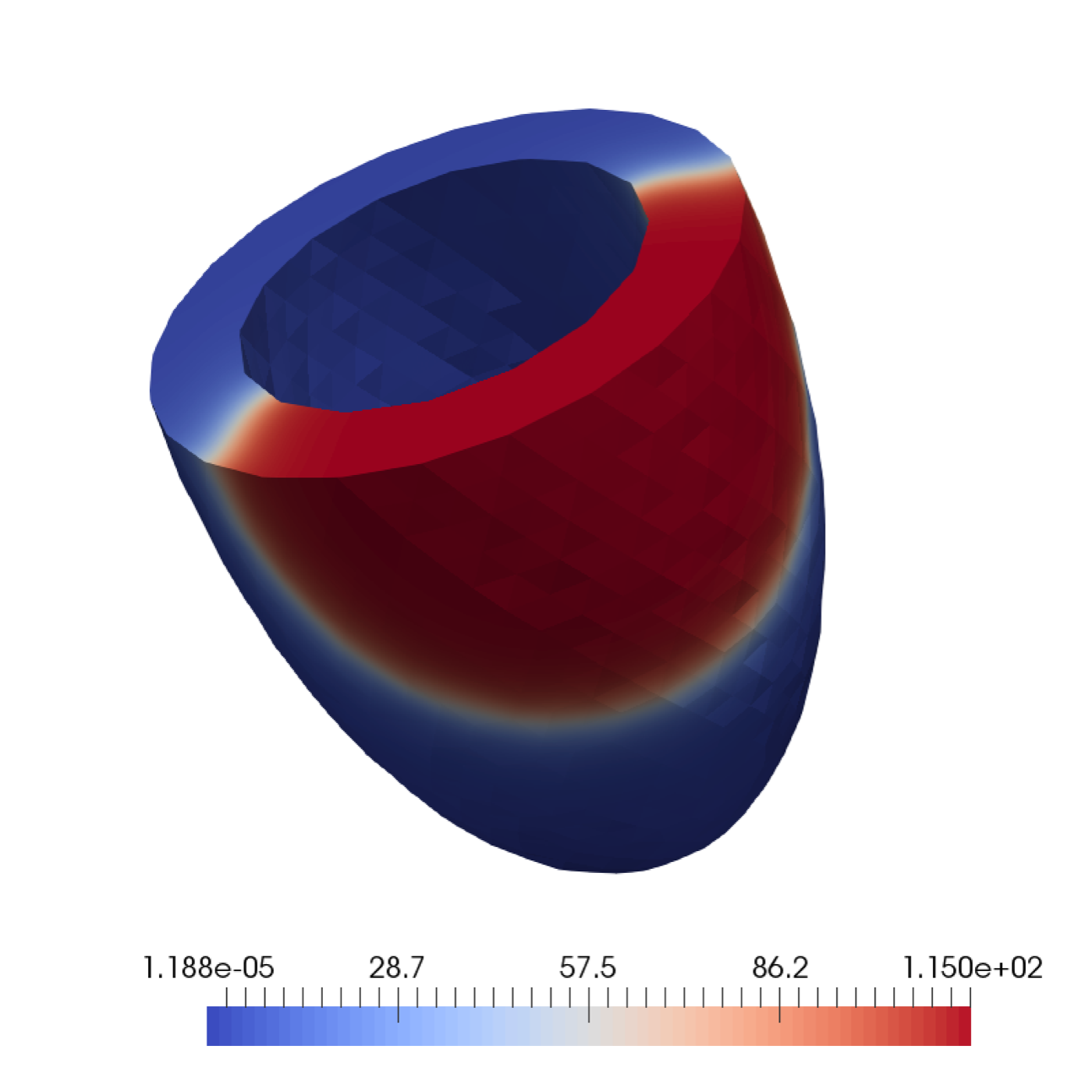}
	            \includegraphics[width=0.4\textwidth]{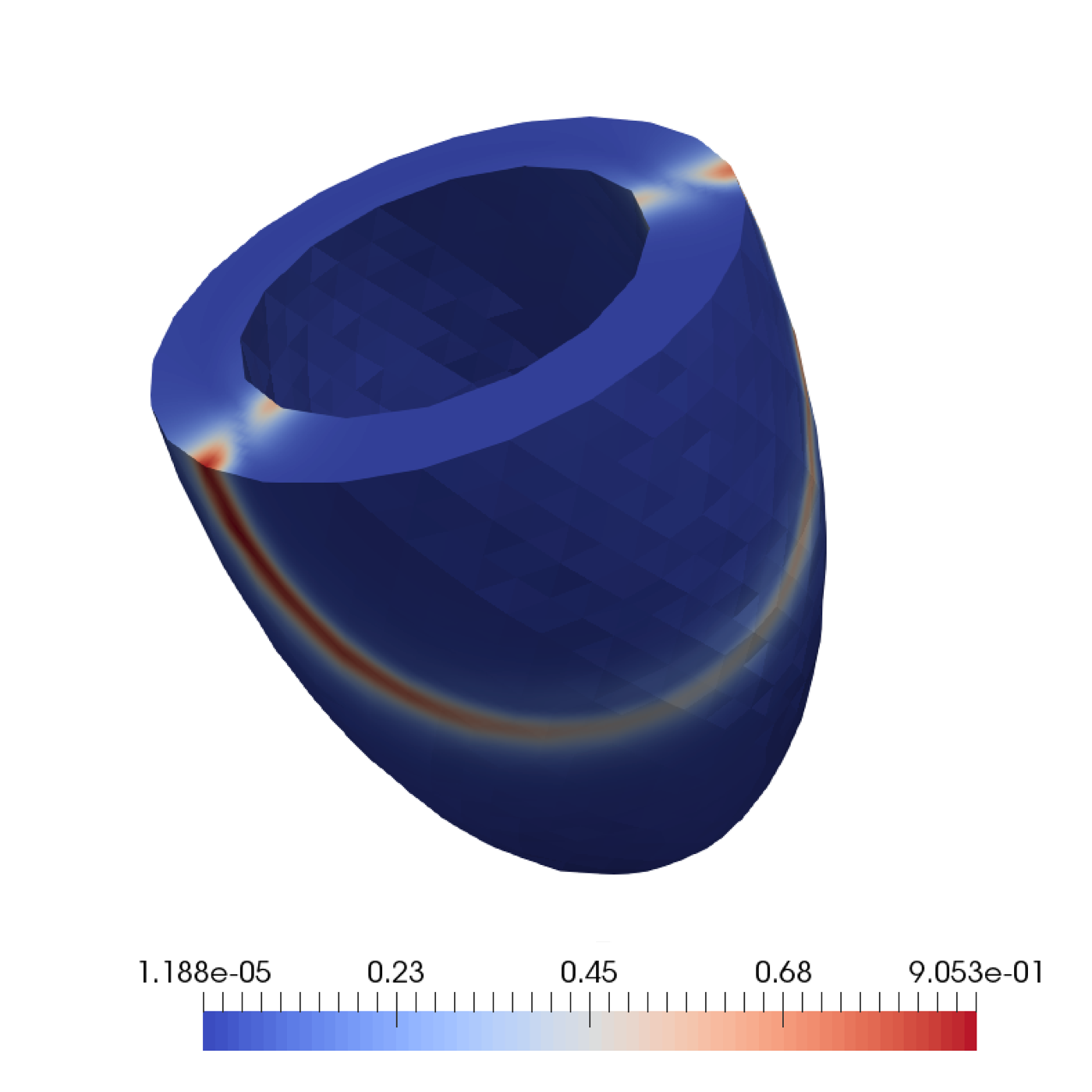}
                 \caption{Mean (left) and variance (right) at the final time state for the ventricle.}
                 \label{fig::MC-ventricle}
                 \end{figure}

  We conduct a Monte-Carlo study of $N=10000$ samples on the fine discretization levels in order to obtain a reference solution. In Figures $\ref{fig::MC-cube}$  and  $\ref{fig::MC-ventricle}$, we show the mean and the variance at the final time state for the cube and the ventricle, respectively. 
  
\begin{figure}[H] 
	            \centering
	            \includegraphics[width=0.45\textwidth]{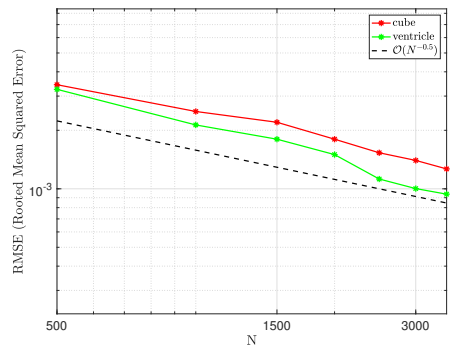}
                 \caption{Convergence rate for the MC estimator for the cube and the ventricle.}
                 \label{fig::MC-convergence}
                 \end{figure}

  In Section $\ref{section:methods}$, we present the theoretical semi-linear convergence of the MC method (cf. Equation $(\ref{eq:stochastic-error}$)). We verify this in Figure $\ref{fig::MC-convergence}$ in which we plot the rooted mean square error (RMSE) against the number of samples ($N$) used for the estimation process. It can be observed how the RMSE indeed behaves as $ \text{RMSE} = \mathcal{O}(N^{-0.5}) $. Notice that these plots rely on averaged multiple runs.
 
\subsubsection{Mutilevel Monte-Carlo in 3D+1}
\begin{figure}[H]
    \centering
      \begin{subfigure}{0.45\textwidth}
        \includegraphics[width=\textwidth]{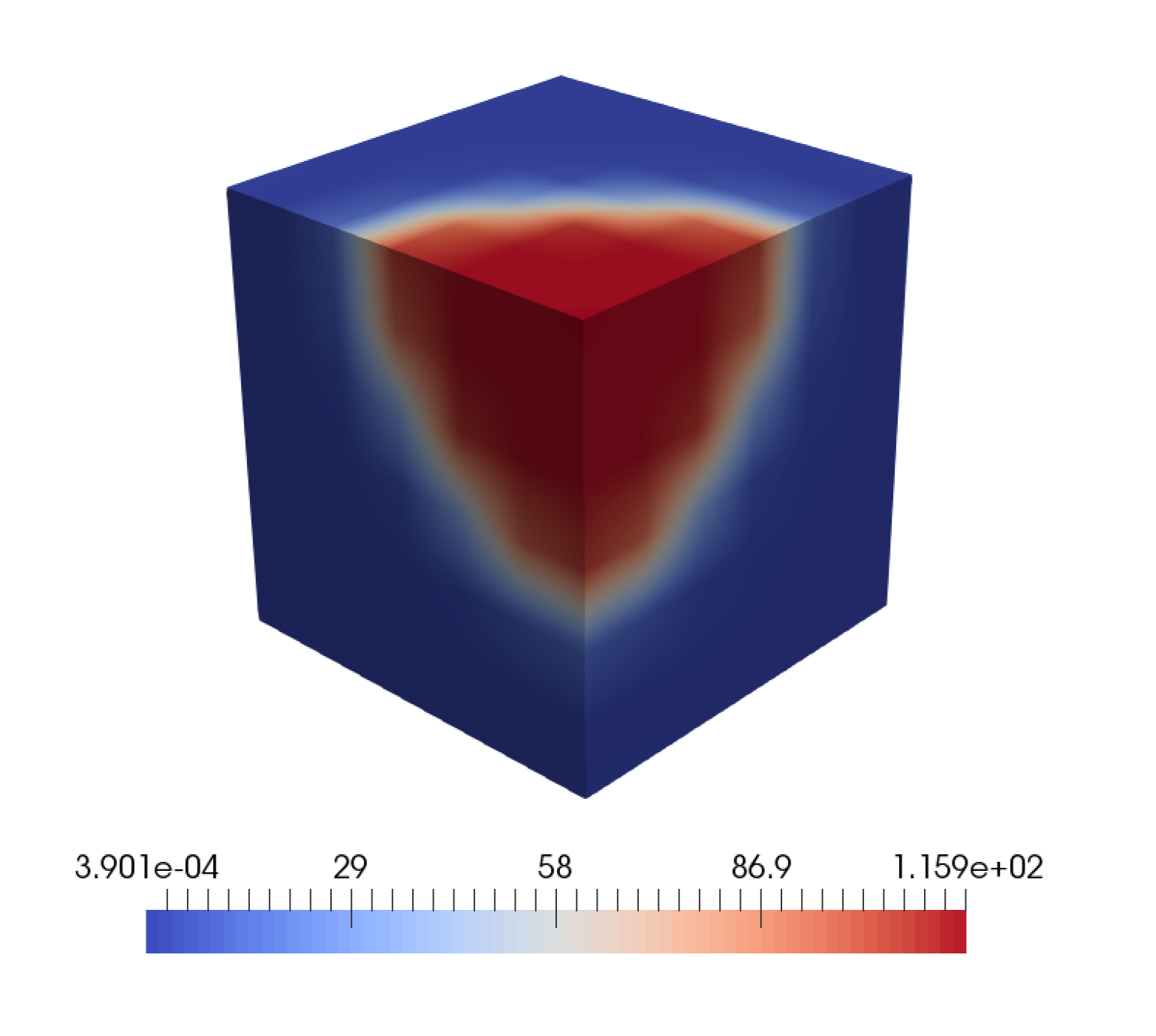}
          \caption{Mean at coarse level.}
          \captionsetup{belowskip=0pt}
         
      \end{subfigure}
      \hfill
      \begin{subfigure}{0.45\textwidth}
        \includegraphics[width=\textwidth]{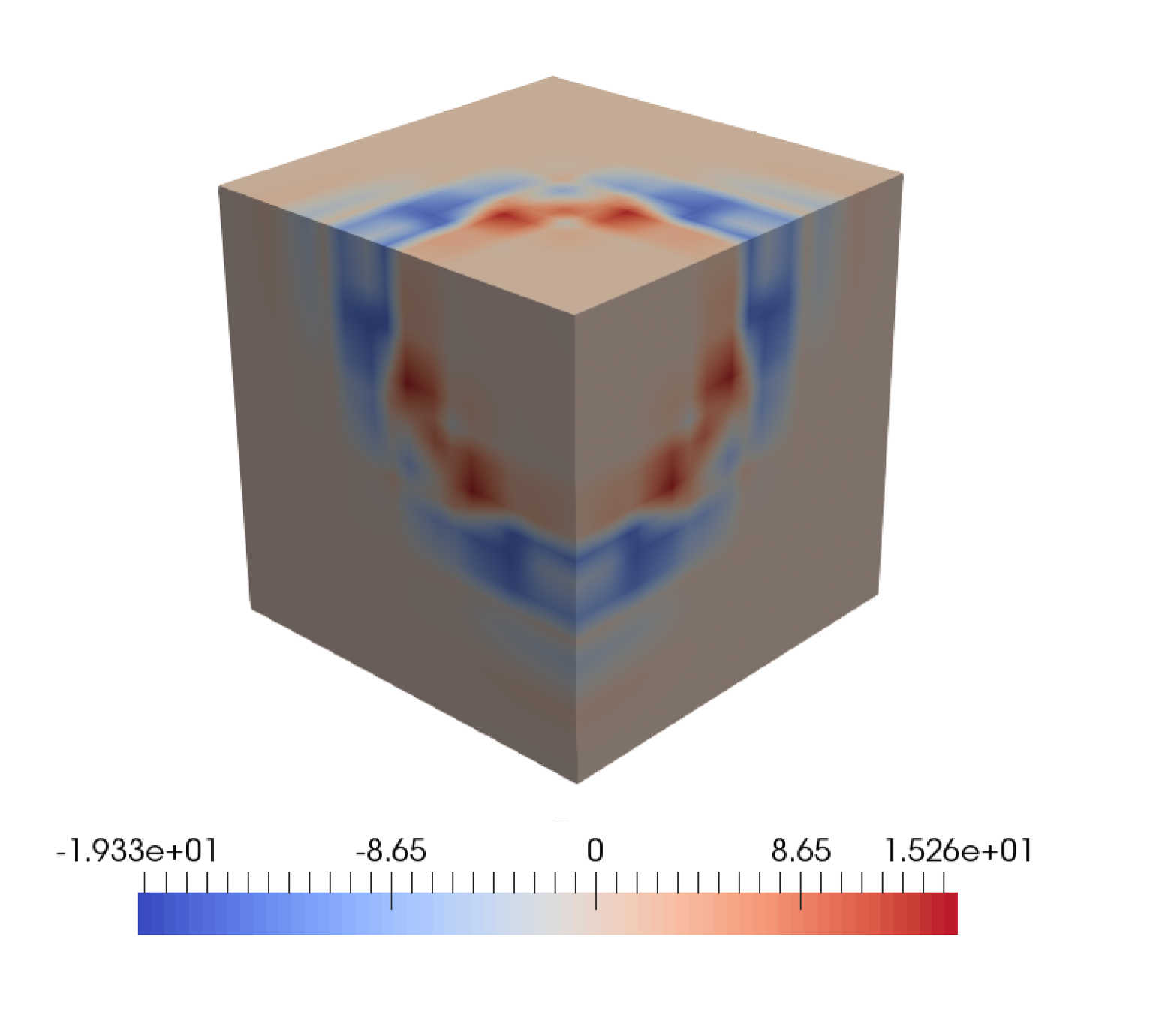}
          \caption{Correction between levels 0 and 1.}
          \captionsetup{belowskip=0pt}
      \end{subfigure}
      
        \begin{subfigure}{0.45\textwidth}
        \includegraphics[width=\textwidth]{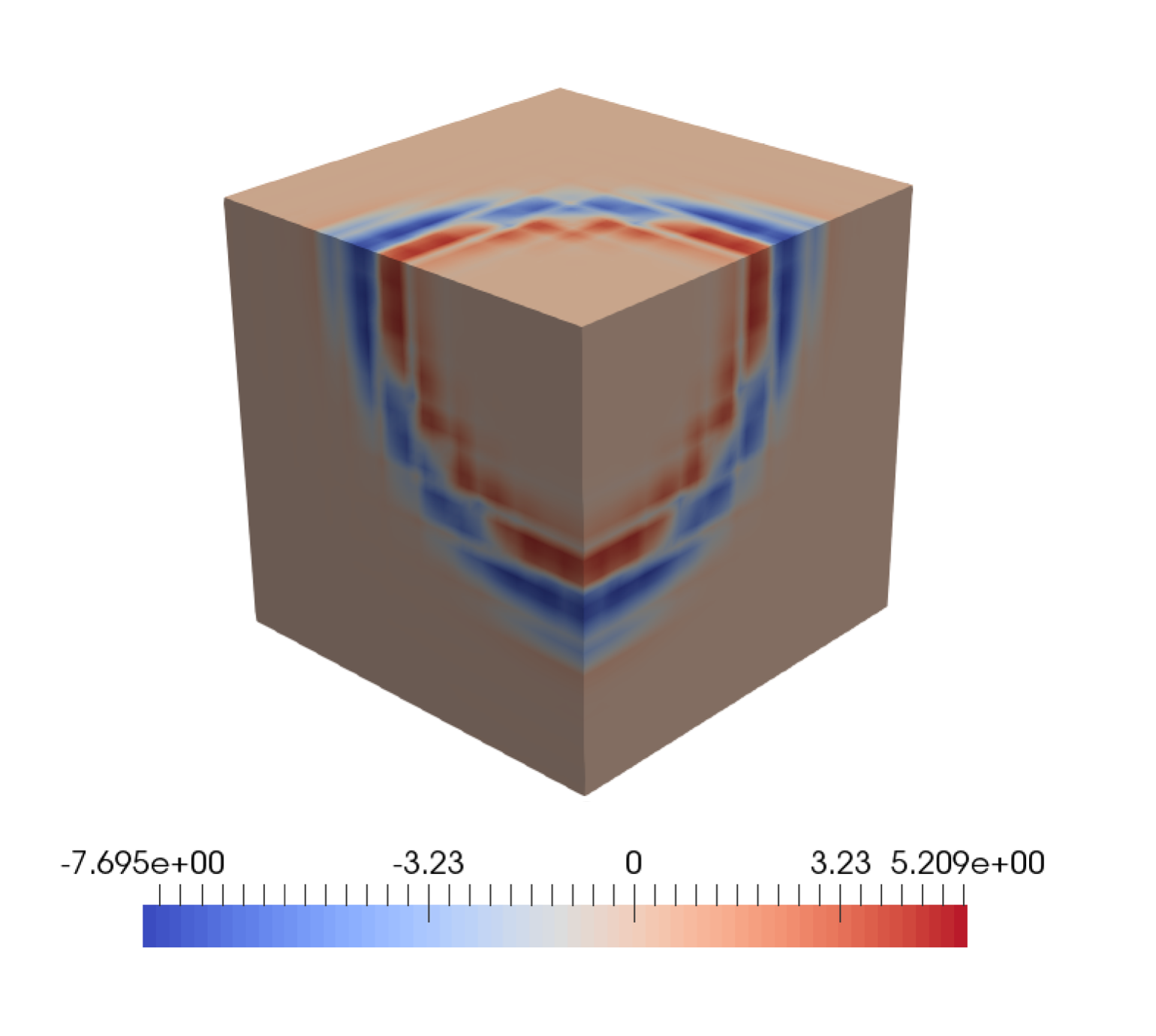}
          \caption{Correction between levels 1 and 2.}
          \captionsetup{belowskip=0pt}
        
      \end{subfigure}
      \hfill
      \begin{subfigure}{0.45\textwidth}
        \includegraphics[width=\textwidth]{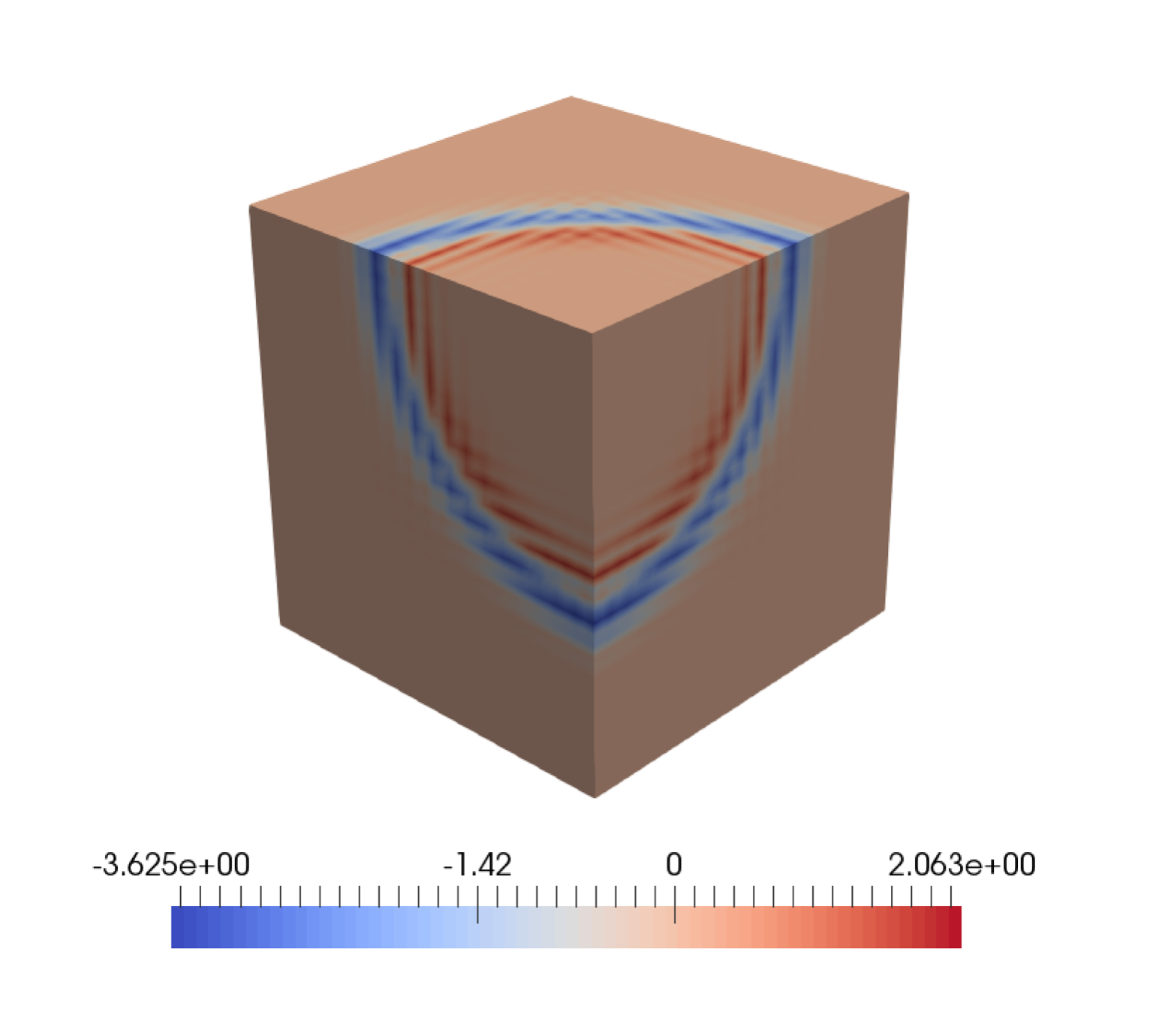}
          \caption{Correction between levels 2 and 3.}
          \captionsetup{belowskip=0pt}
      \end{subfigure}
      
    \caption{Mean at coarse level and corrections between successive levels for the cube.}
    \label{fig::MLMC-estimator-cube}
      
\end{figure}

\begin{figure}[htbp!]
    \centering
      \begin{subfigure}{0.3\textwidth}
      \centering
    
        \includegraphics[width=\textwidth]{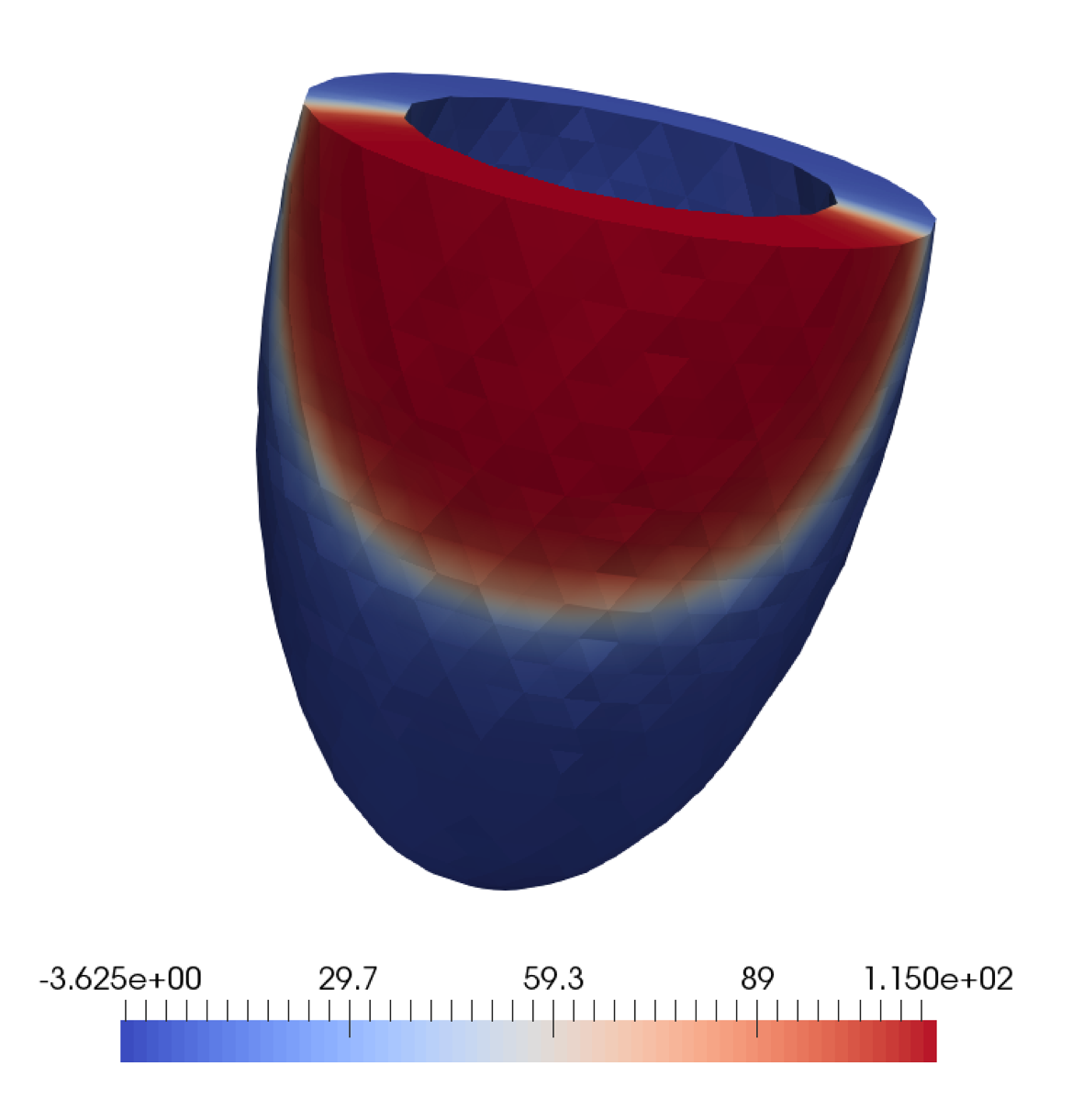}
          \caption{Mean at coarse level.}
          \captionsetup{justification=centering}
         
      \end{subfigure}
      \hfill
      \begin{subfigure}{0.3\textwidth}
      \centering
      \vspace{0.47cm}
        \includegraphics[width=\textwidth]{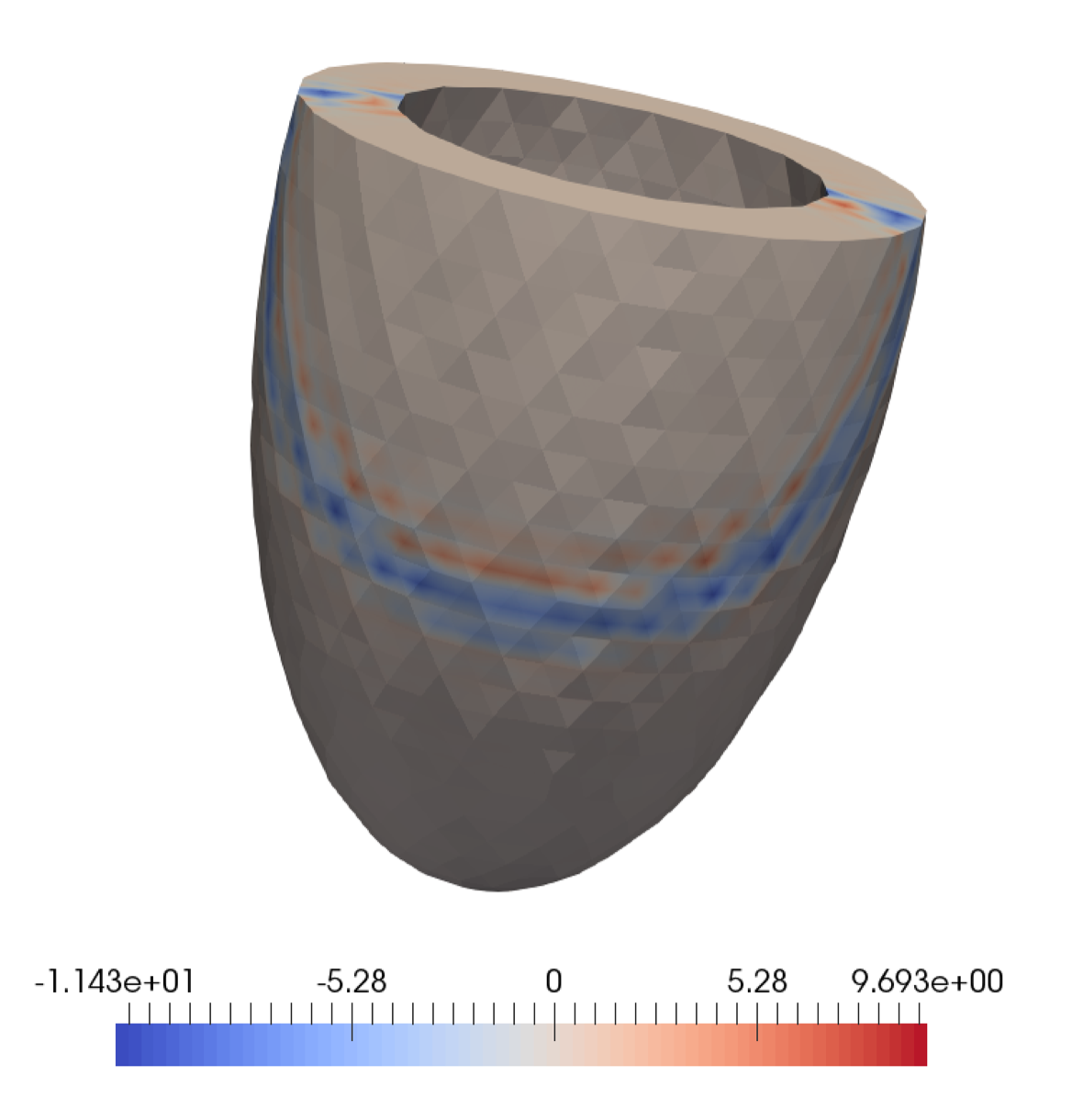}
        \captionsetup{justification=centering}
          \caption{Correction between levels 0 and 1.}
        
      \end{subfigure}
      \hfill
        \begin{subfigure}{0.3\textwidth}
        \vspace{0.47cm}
        \centering
        \includegraphics[width=\textwidth]{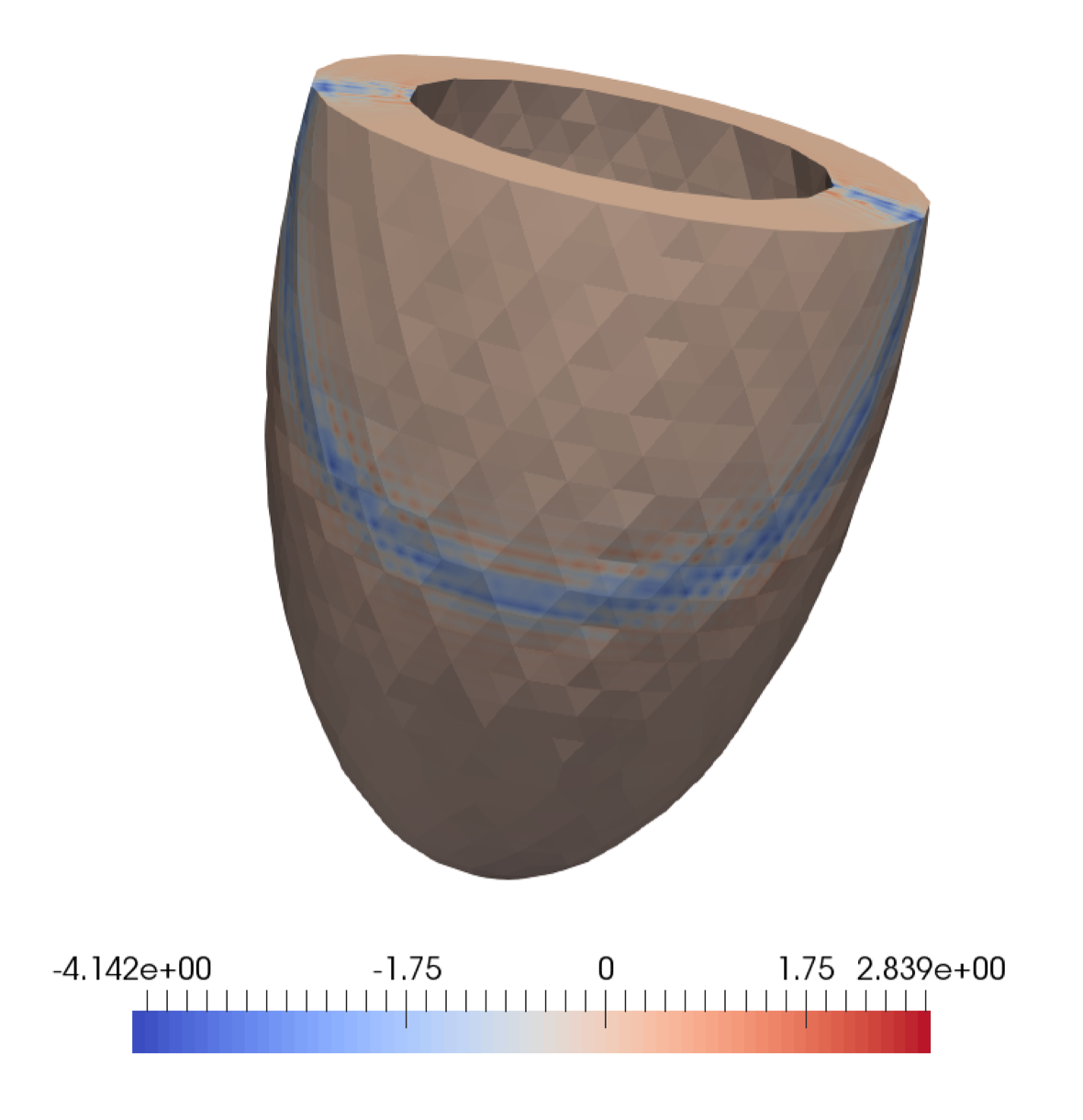}
        \captionsetup{justification=centering}
          \caption{Correction between levels 1 and 2.}
         
      \end{subfigure}
      
    \caption{Mean at coarse level and corrections between successive levels for the ventricle.}
    \label{fig::MLMC-estimator-ventricle}
      
\end{figure}

The setting for the MLMC is very similar to that of the 1D+1 case. We underline the importance of the ideal sampling ratio as detailed at the end of Section $\ref{section:methods}$. We provide in Figures $\ref{fig::MLMC-estimator-cube}$ and $\ref{fig::MLMC-estimator-ventricle}$ a visualization of the MLMC estimator (mean at coarse level and corrections between successive levels) for the cube and the ventricle respectively.

\begin{figure}[h!]
	            \centering
	            \includegraphics[width=0.49\textwidth]{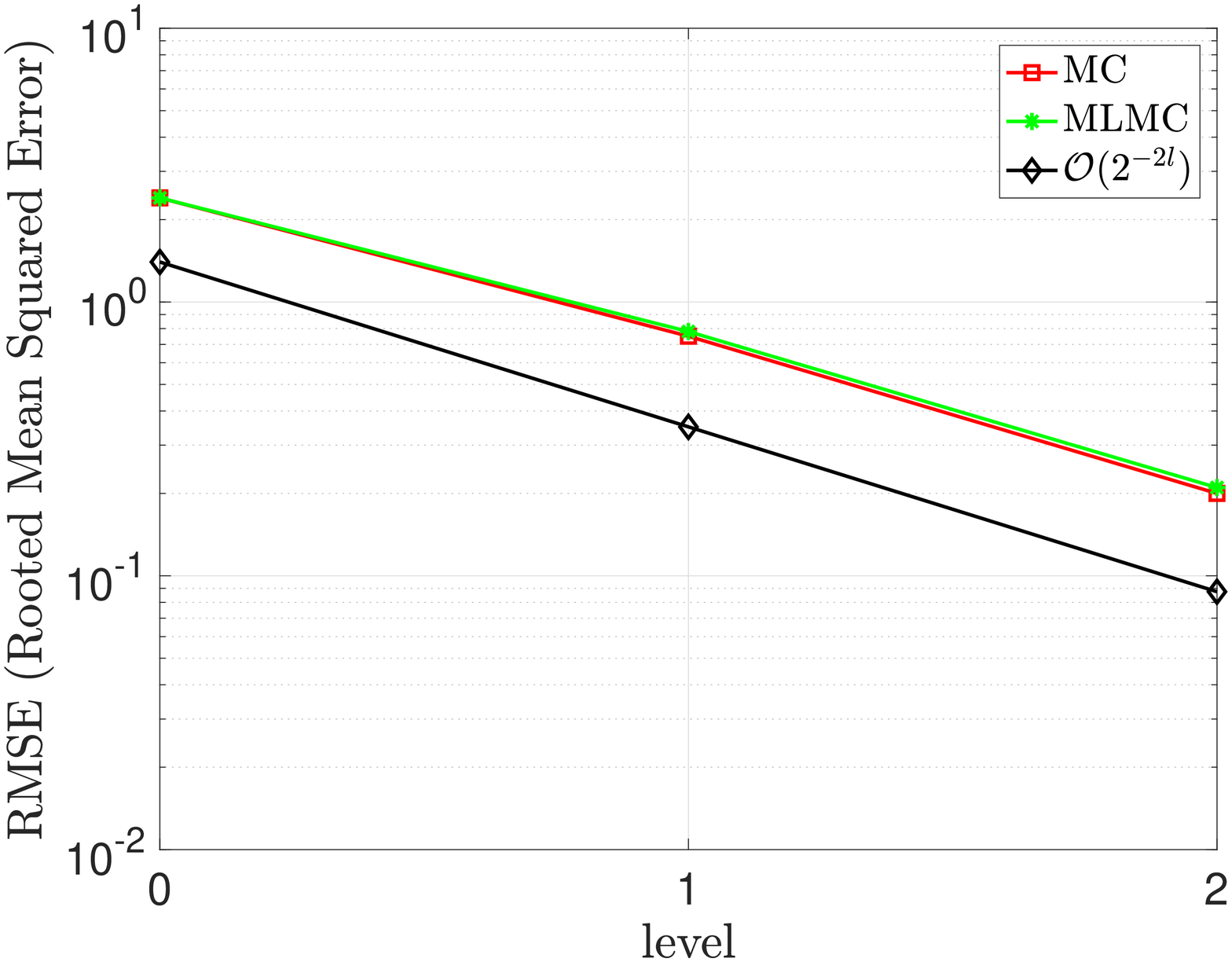}
	            \includegraphics[width=0.49\textwidth]{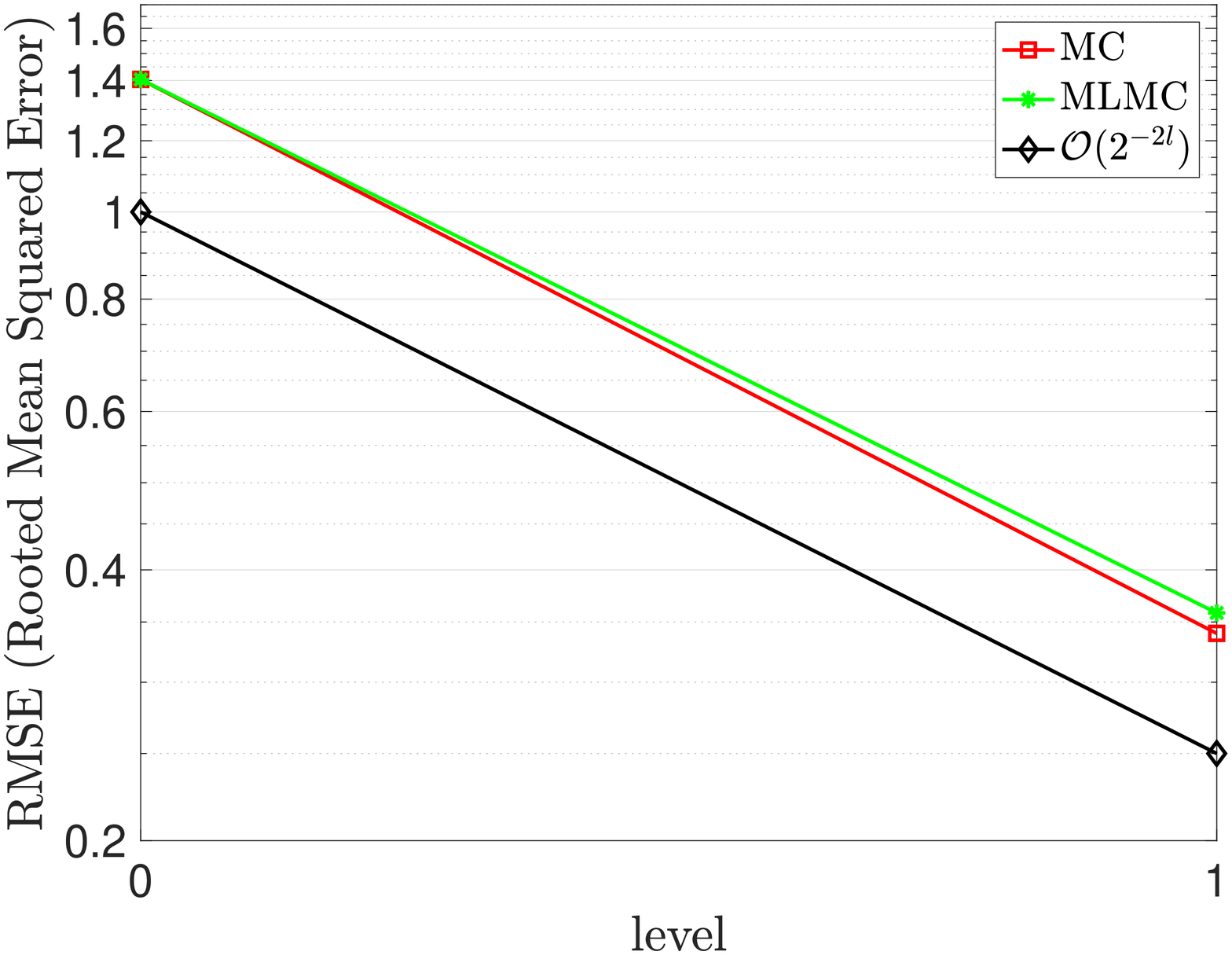}
	            \captionsetup{justification=centering}
                 \caption{Controlled convergence (MC and MLMC) graphs for the cube (left) and the ventricle (right).}
                 \label{fig::Controlled-Convergence}
                 \end{figure}

Figure $\ref{fig::Controlled-Convergence}$ presents the controlled convergence graph of MC and MLMC for both the cube and the ventricle. It demonstrates that the general convergence rate obtained for these two methods is as expected. MC and MLMC have therefore a similar controlled convergence rate.

\begin{figure}[h!]
	            \centering
	            \includegraphics[width=0.49\textwidth]{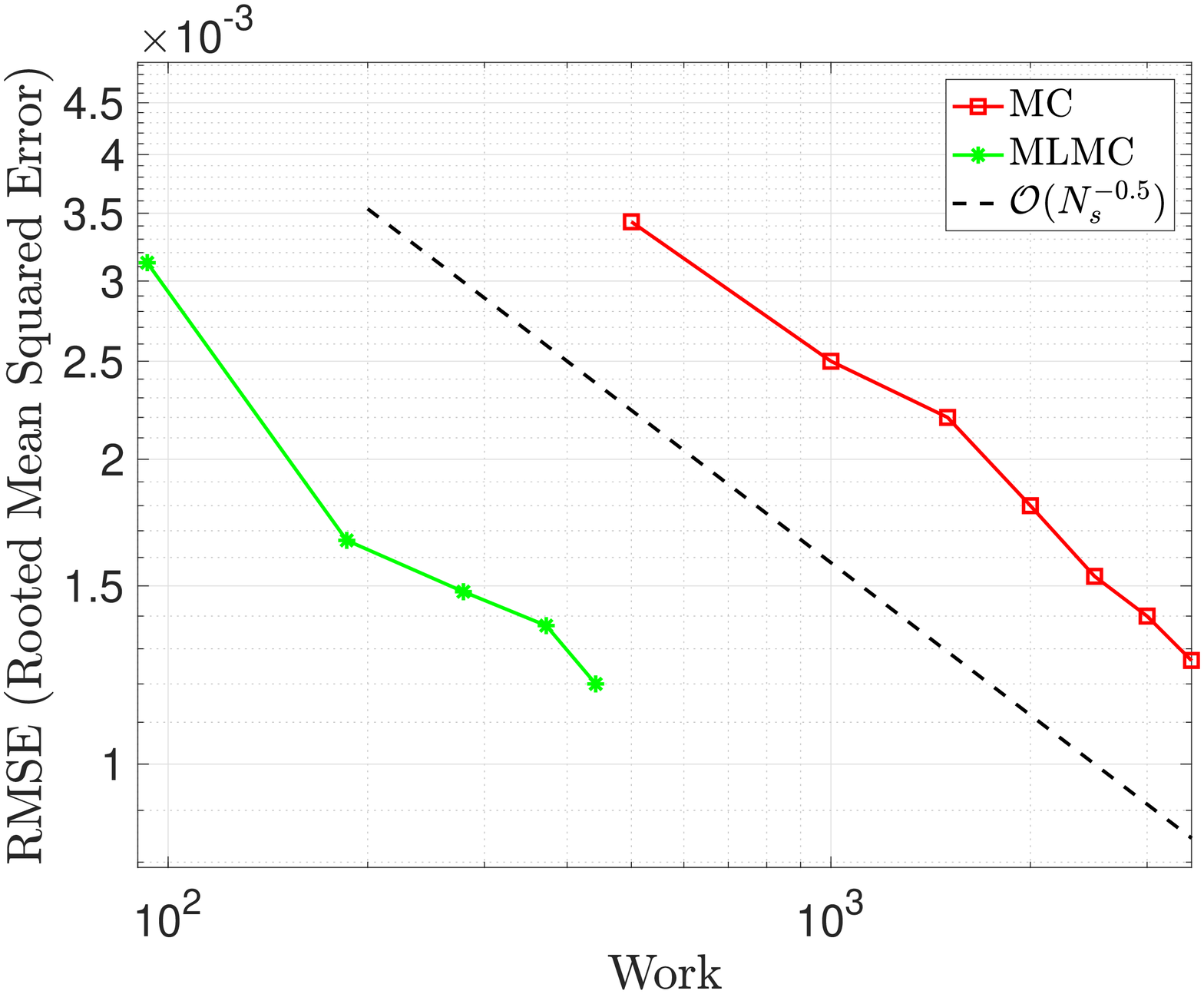}
	            \includegraphics[width=0.49\textwidth]{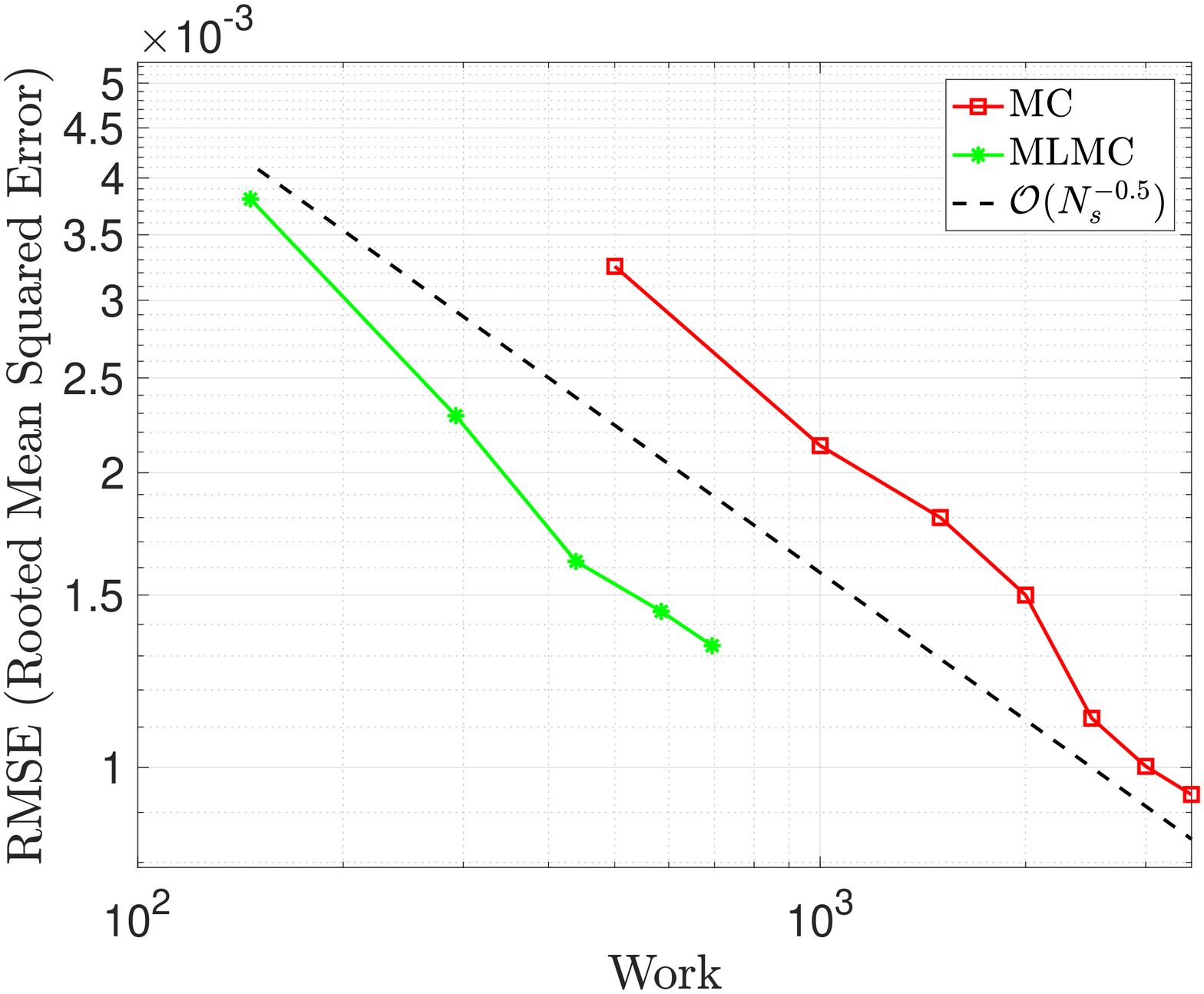}
	            \captionsetup{justification=centering}
                 \caption{Work comparison between MC and MLMC for the cube (left) and the ventricle (right). Works ($W_{\mathcal{MC},L}$ and $W_{\mathcal{MLMC},L}$ as defined in \eqref{eq:work}) are measured on the basis of CPU time.}
                 \label{fig::Work}
                \end{figure}

\paragraph{Work behaviour for a fixed MLMC setting:}
As mentioned in the previous paragraph, the plots for controlled error in Figure $\ref{fig::Controlled-Convergence}$ (or controlled convergence) implicitly indicate that MLMC achieves the same precision with less effort.
 In order to better underline this aspect, we attempt here to evaluate this difference of produced work for both methods, whenever the MLMC setting (fine level, number of levels, mesh hierarchy,...) had been previously set. Let us denote with $W'_l$ the expected work for solving a sample at level $l$, $\forall l = 0,\cdots,L$. We also introduce $W_l$ to be defined as
 \[  W_l = W'_{l-1} + W'_l,  \hspace{0.5cm} \forall l = 0,\cdots,L  \]
 with $W'_{-1} \equiv 0 $. We can therefore formulate the total work expression for both methods in the following way
 \begin{equation} \label{eq:work}
      W_{\mathcal{MC},L} = W'_{L} N_L ,  \hspace{0.5cm} W_{\mathcal{MLMC},L} = \sum_{l=0}^L W_l N_l  ,
 \end{equation}
 where $N_l$ is the number of samples computed on each level $l$, $\forall l = 0,\cdots,L .$ We furthermore can use the sampling ratio between levels for MLMC, which we will denote here with $\beta, $ to rewrite the MLMC total work in function of the number of samples at the fine level exclusively
 \[  W_{\mathcal{MLMC},L} =  \left( \sum_{l=0}^L \beta^{L-l} W_l    \right) N_L . \]
 The latter expression allows us to select different values of the number of samples $N_L$ at the fine level, therefore evaluating the work for different samplings. In Figure $\ref{fig::Work}$, we plot the RMSE versus the total amount of work for both methods. We here define the expected work $W'_l$ for computing a sample at level $l$ to be the total running time for solving a sample at the same level, averaged through solving a hundred of samples. Furthermore, we normalize with respect to the time for solving a sample on the fine level. 
 
\begin{figure}[H] 
	            \centering
	            \includegraphics[width=0.49\textwidth]{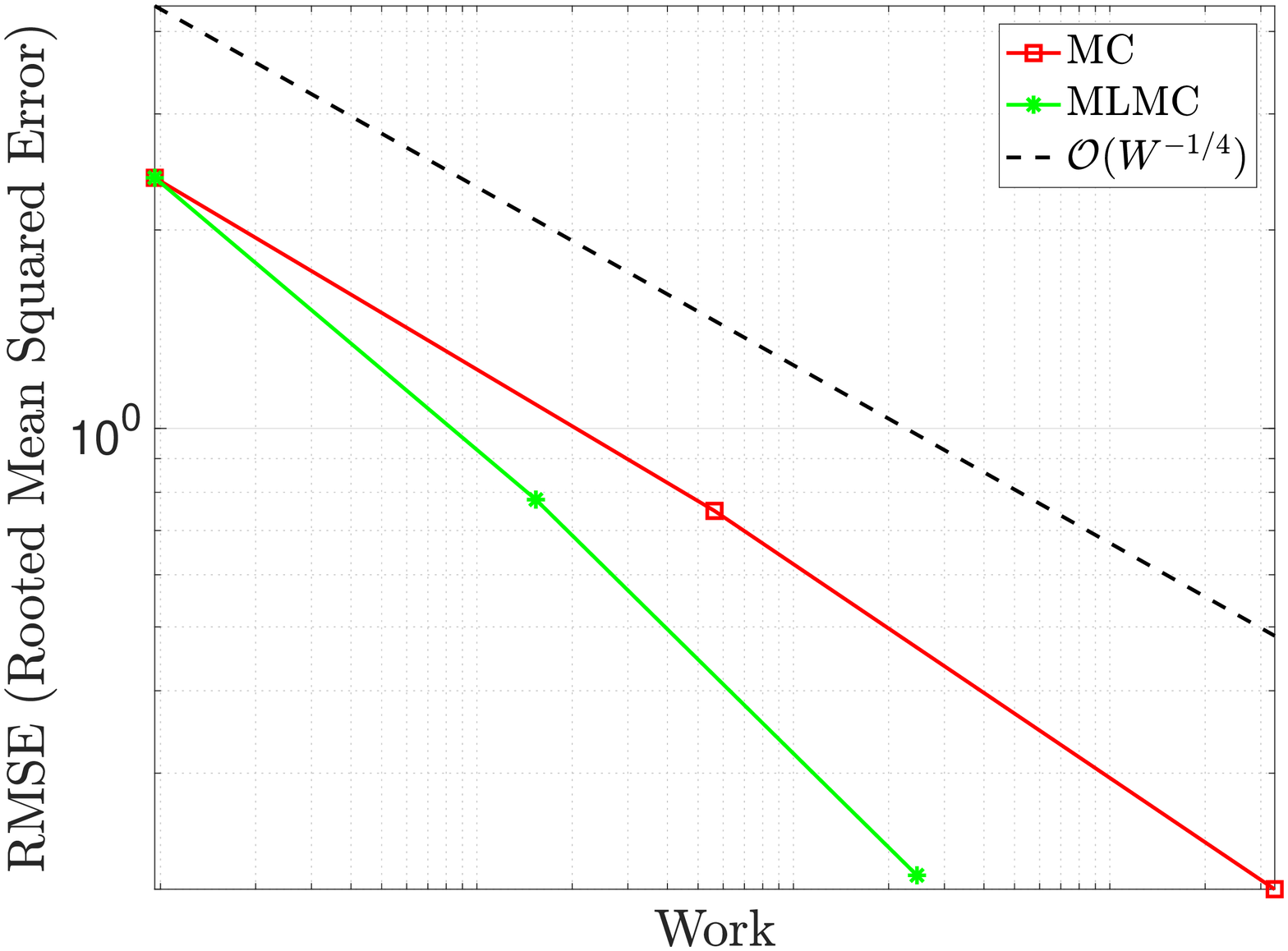}
	            \includegraphics[width=0.49\textwidth]{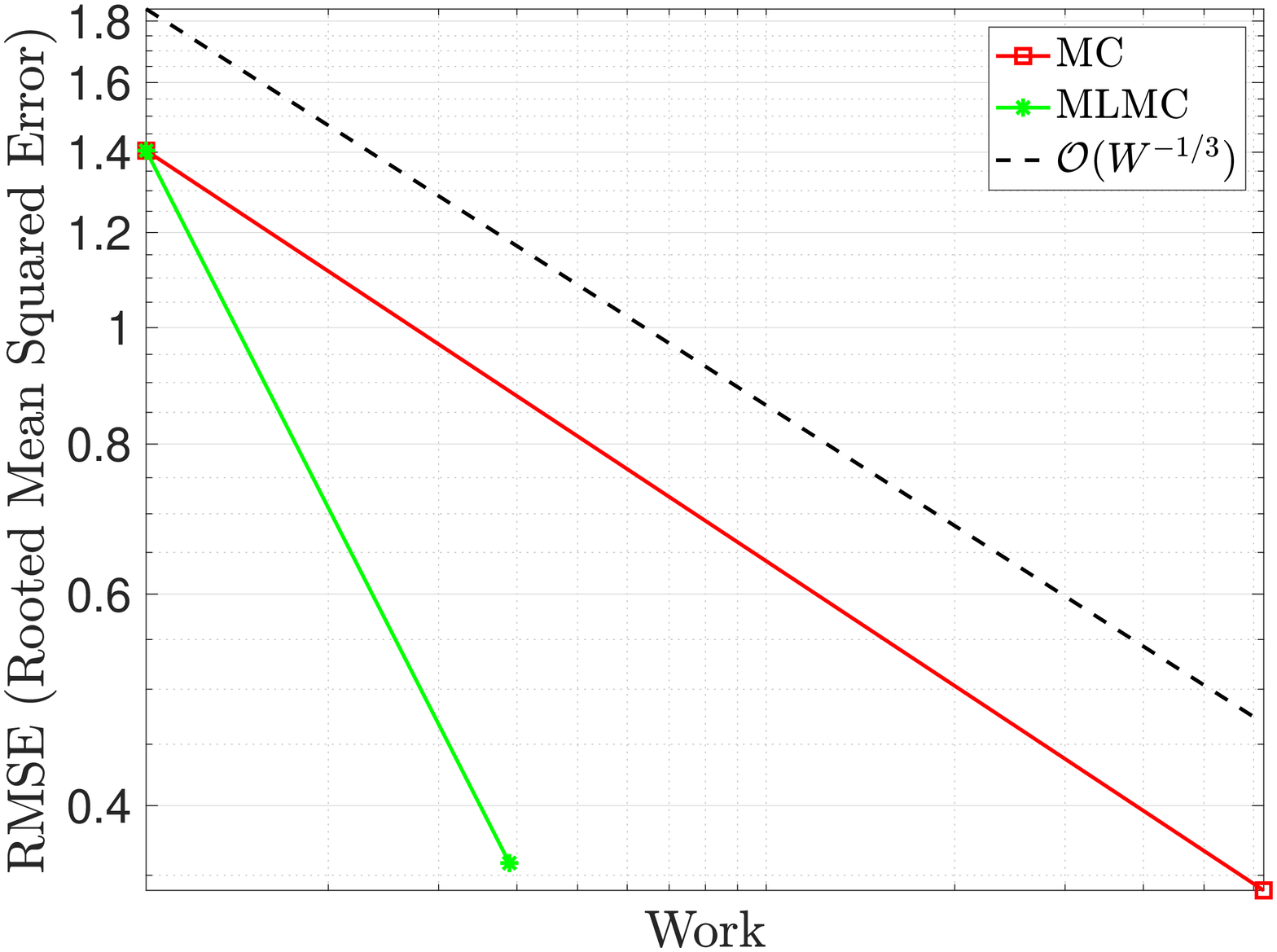}
	            \captionsetup{justification=centering}
                 \caption{MC and MLMC asymptotical work for the cube (left) and the ventricle (right). The "dimensional complexity" ($\gamma d$ as defined in $\eqref{eq::complexity}$) is measured on the basis of CPU time.}
                 \label{fig::Asymptotical-Work}
                \end{figure}

\paragraph{Asymptotical work behaviour of MC/MLMC: }
Equivalently to the controlled convergence concept, we would like now to study the work in the context of the controlled error. 
Following the theory, in order to achieve an error of order $2^{-2l},$ the cost for solving a sample with linear FE is given by 
\[ C_{\text{FE},l} = 2^{\gamma d l} ,\]
where $\gamma$ is the complexity of the solution method used and $d$ is the dimension for the PDE problem considered (equal to $4$ in our case). The number of samples to execute on a level $l$ for getting an error of order $2^{-2l}$ is given by 

\[ N_{\mathcal{MC},l} = 2^{4l}. \]
From these, one can recover the total amount of work required by both methods for a fine discretization level $L$
\begin{equation} \label{eq::Work-MC}
W_{\mathcal{MC},L} = C_{\text{FE},L} N_{\mathcal{MC},L} = 2^{\gamma d L} 2^{4L} = 2^{ (\gamma d + 4) L }  
\end{equation}
and 
\begin{equation}
    W_{\mathcal{MLMC},L} = \sum_{l=0}^{L} C_{\text{FE},L} N_{\mathcal{MC},L-l} = \sum_{l=0}^{L} 2^{\gamma d l} 2^{4(L-l)} =  2^{4L} \sum_{l=0}^{L} 2^{ (\gamma d - 4) l}
\end{equation}
This can be rewritten as 

\begin{equation} \label{eq::Work-MLMC}
     W_{\mathcal{MLMC},L} = 
    \begin{cases}
      L 2^{4L}  \text{  if  }  \gamma d = 4 , \\
      \dfrac{ 2^{\gamma d L} - 2^{4L}  }{2^{\gamma d - 4} - 1} \text{        if  } \gamma d  \neq  4 .
    \end{cases}
\end{equation}
Therefore, in case dimension $d$ and complexity $\gamma$ are known, equations $\eqref{eq::Work-MC}$ and $\eqref{eq::Work-MLMC}$ give a theoretical estimate on the asymptotical behaviour for the work produced by both MC and MLMC. In our case, since we are measuring the work in terms of time to solution (CPU time), we can estimate the product of those values as 
\begin{equation} \label{eq::complexity}
    \gamma d = \log_2  (W'_{l+1}/W'_l) .
\end{equation}
The following estimates yield
\begin{itemize}
    \item $\gamma d \approx 4$ for the cube 
    \item $\gamma d \approx 2$ for the ventricle 
\end{itemize}
The MC and MLMC produced work for the cube experiments are therefore given by
\begin{equation} \label{eq::Work-cube}
    W_{\mathcal{MC},L} = 2^{8L},        \hspace{0.5cm}        W_{\mathcal{MLMC},L} = L 2^{4L}.
\end{equation}
For the ventricle experiments, we obtain
\begin{equation} \label{eq::Work-ventricle}
    W_{\mathcal{MC},L} = 2^{6L},        \hspace{0.5cm}        W_{\mathcal{MLMC},L} =  \dfrac{3(2^{4L} - 2^{2L})}{4}.
\end{equation}
This demonstrates very clearly that the asymptotic behavior of MLMC in terms of computational work is significantly reduced in comparison with standard MC. 
We illustrate this in Figure \ref{fig::Asymptotical-Work}.
We recall that the controlled RMSE $e_l$ at level $l$ is of order $2^{-2l}$. From this, we can deduce that $ e_l^{-0.5} \sim 2^l.$ Putting this together with equations \eqref{eq::Work-cube} and \eqref{eq::Work-ventricle}, shows that for the cube we have
\[e_l =  \mathcal{O}(W_{\mathcal{MC}}^{-1/4}),\]
whereas for the ventricle
\[e_l =  \mathcal{O}(W_{\mathcal{MC}}^{-1/3}).\]
This is illustrated in  Figure \ref{fig::Asymptotical-Work}.

%
%

\section{Conclusions}
We presented a novel approach aimed at performing uncertainty quantification for time-dependent problems, in the presence of high-dimensional input uncertainties and output quantities-of-interest. We assessed the performance of the proposed framework by showing in details its application to cardiac electrophysiology. The results show that the proposed method attains the ideal convergence rates predicted by the theory. Our methodology relies on a close integration of multilevel Monte Carlo methods, parallel iterative solvers, and a space-time discretization. This allows to exploit their synergies, such as for the initialization of Newton's method using the solution of past samples when they become available. The proposed method can also be extended to space-time adaptivity and time-varying domains, both of which are very relevant in the context of cardiac electrophysiology and will be addressed in future works. 

\section{Acknowledgements}
The authors would like to thank the Swiss National Science Foundation (SNSF) for their support through the project "Multilevel Methods and Uncertainty Quantification in Cardiac Electrophysiology" in collaboration with the University of Basel (grant agreement SNSF-205321\_169599), but also through the project "HEARTFUSION: Imaging-driven Patient-specific Cardiac Simulation" in collaboration with the University of Bern (grant agreement SNSF- 169239).
Furthermore, the authors would like to thank the Deutsche Forschungsgemeinschaft (DFG) which has supported parts of the software used here in the SPPEXA program ``EXASOLVERS - Extreme Scale Solvers for Coupled Problems'' and SNSF under the lead agency grant agreement SNSF-162199.  
Part of the software tools utilized in this paper were developed as part of the activities of the Swiss Centre for Competence in Energy Research on the Future Swiss Electrical Infrastructure (SCCER-FURIES), which is financially supported by the Swiss 
Innovation Agency (Innosuisse - SCCER program). The authors gratefully acknowledge financial support by the Theo Rossi di Montelera Foundation, the Mantegazza Foundation, and FIDINAM to the Center of Computational Medicine in Cardiology. 

\bibliography{main}

\end{document}